\pdfoutput=1
\documentclass[cernpreprint,english]{na61doc}

\usepackage{cite}
\usepackage{blindtext}
\usepackage{enumerate}
\usepackage{subcaption}
\usepackage{lineno}
\usepackage{mathtools}
\usepackage{amsmath}
\usepackage{float}
\usepackage{colortbl}
\definecolor{darkred}{rgb}{0.5,0,0}
\definecolor{darkblue}{rgb}{0,0,0.5}
\definecolor{firebrick}{rgb}{0.75,0.125,0.125}
\definecolor{darkgreen}{rgb}{0,0.5,0}

\usepackage[colorlinks=true,linkcolor=firebrick,citecolor=darkgreen,urlcolor=darkblue]{hyperref}

\renewcommand{\NASixtyOne}{\mbox{NA61/SHINE}\xspace}

\newcommand{\pp}{\mbox{\textit{p}+\textit{p}}\xspace}
\newcommand{\pPb}{\mbox{\textit{p}+Pb}\xspace}

\newcommand{\eV}{\ensuremath{\mbox{e\kern-0.1em V}}\xspace}
\newcommand{\GeV}{\ensuremath{\mbox{Ge\kern-0.1em V}}\xspace}
\newcommand{\MeV}{\ensuremath{\mbox{Me\kern-0.1em V}}\xspace}
\newcommand{\GeVc}{\ensuremath{\mbox{Ge\kern-0.1em V}\!/\!c}\xspace}
\newcommand{\GeVcc}{\ensuremath{\mbox{Ge\kern-0.1em V}\!/\!c^2}\xspace}
\newcommand{\AGeV}{\ensuremath{A\,\mbox{Ge\kern-0.1em V}}\xspace}
\newcommand{\AGeVc}{\ensuremath{A\,\mbox{Ge\kern-0.1em V}\!/\!c}\xspace}
\newcommand{\MeVc}{\ensuremath{\mbox{Me\kern-0.1em V}/c}\xspace}

\newcommand{\dd}{\ensuremath{{\textrm d}}\xspace}
\newcommand{\dedx}{\ensuremath{\dd E\!/\!\dd x}\xspace}

\newcommand{\mt}{\ensuremath{m_{\textnormal T}}\xspace}
\newcommand{\kt}{\ensuremath{K_{\textnormal T}}\xspace}

\newcommand{\CernVM}{\textsc{Cern\-\kern-0.05emVM}\xspace}

\ShineTitle{Two-pion femtoscopic correlations in Be+Be collisions at $\sqrt{s_{\textnormal{NN}}} = 16.84$ GeV measured by the \NASixtyOne at CERN}

\PreprintIdNumber{CERN-EP-2023-013}

\ShineJournalRef{Eur.Phys.J.C 83 (2023) 10, 919}
\ShineDOI{10.1140/epjc/s10052-023-11997-8}
\ShineAbstract{This paper reports measurements of two-pion femtoscopic correlations in Be+Be collisions at a beam momentum of 150\AGeVc (energy available in the center-of-mass system for nucleon pair  $\sqrt{s_{\textnormal{NN}}} = 16.84$ GeV) by the \NASixtyOne experiment at the CERN SPS accelerator. The obtained momentum space correlation functions can be well described by a L\'evy distributed source model. The transverse mass dependence of the L\'evy source parameters is presented, and their possible theoretical interpretations are discussed. The results show that the Lévy exponent $\alpha$ is approximately constant as a function of \mt , and far from both the Gaussian case of $\alpha = 2$ or the conjectured value at the critical endpoint, $\alpha = 0.5$. The radius scale parameter $R$ shows a slight decrease in \mt , which can be explained as a signature of transverse flow. Finally, an approximately constant trend of the intercept parameter $\lambda$ as a function of \mt was observed, similar to previous NA44 S+Pb results (obtained with a Gaussian approximation, but unlike RHIC results).}
\begin{document}
\maketitle

%***********************************************************************************
\section{Introduction}
This paper reports measurements of quantum-statistical femtoscopic correlation functions for identified, like-sign charged pion pairs produced in central Be+Be collisions at 150\textit{A} GeV/\textit{c} beam momentum.

The method of quantum-statistical (Bose-Einstein) correlations was first applied in astrophysical intensity correlation measurements by R. Hanbury Brown and R. Q. Twiss (HBT)~\cite{HanburyBrown:1956bqd} in order to determine the apparent angular diameter of stellar objects. Later, a similar quantum-statistical method was applied in momentum correlation measurements for proton-antiproton collisions~\cite{Goldhaber:1959mj,Goldhaber:1960sf} to obtain information on the size parameters of particle emission sources in high-energy particle collisions. Since then, quantum-statistical (HBT) correlation measurements have become a standard tool for experimental characterization of the probability density in particle the emission process, i.e.\ the source function, which sheds light on the spatio-temporal structure of particle emission. This experimental method also largely contributed to the understanding of the hydro-dynamical nature of the produced strongly interacting matter. In fact, the pair momentum dependence of Gaussian shaped source radii~\cite{Adler:2004rq,Afanasiev:2009ii} can be well explained by a hydro-dynamical expansion. The shape of the particle emitting source was furthermore suggested to be affected by the nature of the quark-hadron transition~\cite{Csorgo:2005it}. Hence, exploring HBT correlations is of utmost importance in the quest for understanding the nature of the matter created in relativistic heavy-ion collisions.

The results presented in this paper were obtained by the \NASixtyOne~\cite{Abgrall:2014xwa} experiment at the CERN Super Proton Synchrotron accelerator. This measurement is part of the \NASixtyOne strong interaction program investigating the properties of the onset of deconfinement and searching for the possible existence of the critical point of strongly interacting matter. This goal is pursued by the \NASixtyOne Collaboration through a beam energy scan with various nucleus-nucleus collisions. This strategy allows to systematically investigate properties of the phase diagram of strongly interacting matter~\cite{Antoniou:2006mh}. 

Within the framework of statistical models, the data on particle production suggest that with increasing collision energy, the temperature increases and the baryon chemical potential of strongly interacting matter decreases at freeze-out, whereas by increasing the nuclear mass number of the colliding nuclei, the temperature decreases~\cite{Becattini:2005xt,Vovchenko:2015idt}.
As a result of the \NASixtyOne research program, a large set of collision data on \pp, \pPb, Be+Be, Ar+Sc, Xe+La, and Pb+Pb collision systems has already been recorded. An upgrade of the \NASixtyOne experiment was completed in 2022, and further high-statistics data on Pb+Pb collisions will be collected in the near future~\cite{PbAddendum}. A basic reference has already been established with \pp, Be+Be, and Ar+Sc interactions on particle spectra and multiplicities~\cite{Abgrall:2013pp_pim,NA61SHINE:2017fne,NA61SHINE:2019xkb, NA61SHINE:2021nye,NA61SHINE:2020czq, NA61SHINE:2020ggt}. The present paper provides results on Bose-Einstein correlations of identified, like-sign pion pairs in 0--20\% centrality selected $^7$Be+$^9$Be collisions. The data were recorded in 2011, 2012, and 2013 using a secondary $^7$Be beam produced by fragmentation of the primary Pb beam from the CERN SPS~\cite{Gazdzicki:1322135}.

The paper is organized as follows. Section~\ref{sec:BEIntro} recalls the fundamental theory behind the technique of Bose-Einstein correlations in order to fix notations. In Section~\ref{sec:NA61SHINE}, the \NASixtyOne detector is described. In Section~\ref{sec:Analysis}, the details of the analysis procedure are discussed. In Section~\ref{sec:Results}, the obtained experimental results are presented. The paper closes with Section~\ref{sec:Conclusion}, summarizing the results and conclusions.

%***********************************************************************************

\section{Bose-Einstein correlations}
\label{sec:BEIntro}

\subsection{Bose-Einstein correlation functions}
The two-particle Bose-Einstein correlations are defined in terms of the single- and the two-particle invariant momentum distributions $N_1$ and $N_2$ as:

\begin{equation}
    C_2(p_1,p_2) = \frac{N_2(p_1,p_2)}{N_1(p_1)N_1(p_2)},
    \label{e:twopartcorr}
\end{equation}
where $p_1$ and $p_2$ are the momenta of the individual particles. If the $S(x, p)$ (source function) denotes the probability density of particle creation at space-time point $x$ and momentum $p$, the momentum distribution of emitted particles can be expressed via this source function as~\cite{Pratt:1990zq}:

\begin{align}
N_1(p_1) &= \int S(x_1,p_1) |\Psi_p(x_1)|^2d^4x_1, \\
N_2(p_1,p_2) &= \int S(x_1,p_1)S(x_2,p_2)|\Psi_{p_1,p_2}(x_1,x_2)|^2 d^4 x_2 d^4x_1,
\end{align}
where $\Psi_p(x)$ and $\Psi_{p_1,p_2}(x_1,x_2)$ are the single- and two-particle wave functions. In the case of the single-particle wave function, $|\Psi_p(x)|^{2} = 1$ holds, whereas for the two-particle wave function, taking into account the Bose-Einstein symmetrisation, one has~\cite{Csorgo:1999sj}:

\begin{equation}
|\Psi_{p_1,p_2}(x_1,x_2)|^2 = 1 + \langle \cos(QX) \rangle,
\end{equation}
where $X = x_1 - x_2$ is the relative coordinate and $Q = p_1-p_2$ is the relative momentum of the pair. In the $QX$ term, a division by $\hbar$ is suppressed, and throughout this paper, we will utilize the $\hbar=1$ convention. Substituting the above equation into Eq.~\eqref{e:twopartcorr}, one infers

\begin{equation}
C_2(p_1,p_2) = 1 + \frac{\tilde{S}(Q,p_1)\tilde{S}(Q,p_2)^*}{\tilde{S}(Q=0,p_1)\tilde{S}(Q=0,p_2)^*},\label{e:modiftwopart}
\end{equation}
where $\tilde{S}$ is the Fourier transform of $S$ in its first variable. If relative momentum $Q$ is much smaller compared to the average momentum of the pair $K = (p_1 + p_2)/2$, then Eq.~\eqref{e:modiftwopart} can be expressed as:

\begin{align}
C_2(Q,K) = 1 + \frac{|\tilde{S}(Q,K)|^2}{|\tilde{S}(Q = 0,K)|^2}.\label{e:ktwopart}
\end{align}
Alternatively, with a simplified notation where the $K$-dependence is suppressed and a normalized source is assumed, one may write $C_2(Q) = 1 + |\tilde{S}(Q)|^2$. This choice is motivated by the so-called smoothness approximation~\cite{Lisa:2005dd}. The dependence on relative momentum $Q$ is stronger than on the average momentum of the pair $K$, hence $Q$ is considered as the more important variable of the correlation function, and the other variable is mostly suppressed in the notation. When the correlation function is parameterized based on an ansatz for the source function, its parameters can depend on $K$. In order to explore the transverse dynamics of the source, the average transverse momentum of the pair, $\kt = \sqrt{K^2_x + K_y^2}$ is introduced, where $K_x$ and $K_y$ are transverse components of $K$. Furthermore, motivated by hydro-dynamical considerations, the dependence on the transverse mass  $\mt=\sqrt{m^2c^4+\kt^2c^2}$ is often studied, where $m$ is the particle mass. The \mt-dependence of the source parameters, such as its width, the so-called HBT scale, or radius $R$, was crucial in understanding the transverse expansion dynamics of the strongly interacting matter~\cite{PHENIX:2004yan,Bekele:2007ee}. One of the main goals of HBT correlation measurements is to estimate the size (or, rather, the correlation length) of the hadron emitting source.

\subsection{Core-halo model}\label{s:corehalo}
Equation~\eqref{e:ktwopart} implies that the correlation function takes the value of 2 at zero relative momentum, or equivalently, if the notation $C_2(Q \rightarrow 0) = 1 + \lambda$ were used, then $\lambda=1$ would follow. However, the intercept parameter $\lambda$ is often smaller than one in experimentally measured correlation functions. The widely accepted explanation for this phenomenon is the core-halo  model~\cite{Bolz:1992hc,Csorgo:1994in}, namely that some correlated particles are produced in decays of long-lived resonances, creating a spatially extended component of the source, their momentum difference being unresolvable by the detector. The core-halo model treats these as belonging to the halo component of the source, while the primordial particles and the decay products of short-lived resonances represent the core. While the latter has a size of a few fm, the former may extend to thousands of femtometers, due to long-lived resonances. One can then break up the source $S$ into $S_{\textnormal{core}}$ and  to $S_{\textnormal{halo}}$ as follows:
\begin{equation}
\tilde{S}(Q,K) = \tilde{S}_{\textnormal{core}}(Q,K) + \tilde{S}_{\textnormal{halo}}(Q,K).
\end{equation}
In experimental measurements, the accessible range of $Q$ values is not smaller than a few \MeVc, due to the finite two-track resolution of the tracking detectors. Because of the large radius of the halo, in the accessible $Q$-range $\tilde{S}_{\textnormal{halo}}(Q,K) \approx 0$ thus, $\tilde{S}(Q,K) \approx \tilde{S}_{\textnormal{core}}(Q,K)$. Given that the Fourier-transform of each of the source components at $Q=0$ equals to the number ($N$) of particles in that component,
\begin{equation}
\tilde{S}_{\textnormal{core}}(0,K) = N_{\textnormal{core}},
\quad \tilde{S}_{\textnormal{halo}}(0,K) = N_{\textnormal{halo}},\quad
\tilde{S}(0,K) = N_{\textnormal{core}} + N_{\textnormal{halo}},
\end{equation}
follows, and therefore one obtains
\begin{equation}
C_2(Q) = 1 + \frac{|\tilde{S}_{\textnormal{core}}(Q)+\tilde{S}_{\textnormal{halo}}(Q)|^2}{|\tilde{S}_{\textnormal{core}}(0)+\tilde{S}_{\textnormal{halo}}(0)|^2} \approx
1 + \frac{|\tilde{S}_{\textnormal{core}}(Q)|^2}{|\tilde{S}_{\textnormal{core}}(0)+\tilde{S}_{\textnormal{halo}}(0)|^2} =
1 + \lambda\frac{|\tilde{S}_{\textnormal{core}}(Q)|^2}{|\tilde{S}_{\textnormal{core}}(0)|^2},
\end{equation}
with
\begin{equation}\label{e:corr_str}
\lambda = \frac{|\tilde{S}_{\textnormal{core}}(0)|^2}{|\tilde{S}_{\textnormal{core}}(0)+\tilde{S}_{\textnormal{halo}}(0)|^2} = \left(\frac{N_{\textnormal{core}}}{N_{\textnormal{core}} + N_{\textnormal{halo}}}\right)^2,
\end{equation}
for the experimentally resolvable $Q$-range. 
Although the core-halo model provides a natural explanation for the phenomenon $C_2(Q \rightarrow 0)<2$, i.e.\ $\lambda<1$ in experimental data, it is important to note that $\lambda\neq 1$ can also be explained by other effects, such as coherent pion production~\cite{Csorgo:1999sj,Weiner:1999th} or background from improperly reconstructed particles. It is evident, however, that measuring $\lambda$ is an important tool in understanding particle creation in relativistic heavy-ion collisions. Other effects, such as Coulomb ones, are discussed more in detail in Sec. \ref{subsec:coulomb} and strong interactions are negligible~\cite{Kincses:2019rug}.

\subsection{L\'evy shaped sources and the QCD critical endpoint}
When the source size (i.e., the HBT scale parameter $R$) or the correlation strength (i.e., the intercept parameter $\lambda$) of the Bose-Einstein correlation is to be measured, a full three-dimensional source reconstruction can be performed if the available statistics allow it. While in our analysis it cannot be proven experimentally that the source is fully spherical, studies of 2D correlation functions in the transverse and longitudinal momentum difference variable show no need to go beyond the 1D approximation. Nevertheless, for a more complete study with increased statistics, utilization of spherical harmonics decomposition~\cite{Kisiel:2009iw} or performing further studies in three dimensions may be adequate. Alternatively, a parametric ansatz for the source shape may be used, and its derived correlation function is fitted to the data in order to determine its shape parameters.  Quite naturally, Gaussian sources lead to Gaussian correlation functions. In the present analysis, a more general ansatz is used, i.e. that of L\'evy shaped sources~\cite{Csorgo:2003uv,Metzler:1999zz}, exhibiting possible power-law tails and also incorporating the Gaussian limit. Correlation functions based on this ansatz have been shown to describe LEP~\cite{L3:2011kzb}, RHIC~\cite{PHENIX:2017ino}, and LHC~\cite{CMS:2017mdg,CMS:2023xyd} data as well.

The spherically symmetric L\'evy distribution is defined as
\begin{equation}
\mathcal{L}(\alpha,R,r)=\frac{1}{(2\pi)^3} \int d^3\zeta e^{i\zeta r} e^{-\frac{1}{2}|\zeta R|^{\alpha}},
\end{equation}
where the parameters of this distribution are $\alpha$ and \textit{R}, the L\'evy stability index and L\'evy scale parameter, respectively, while $r$ is the vector of spatial coordinates and the vector $\zeta$ represents the integration variable. In the case of $\alpha=2$, one recovers the Gaussian distribution, while $\alpha = 1$ is equivalent to the Cauchy distribution, and for $\alpha < 2$, the L\'evy distribution exhibits a power-law tail. Hence determining the parameter $\alpha$ by a fit to experimental data yields a way to estimate the deviation of the source from a Gaussian or a Cauchy shape.

Ideally, the correlation function $C_2$ is investigated as a function of momentum difference in the entire three dimensions, but in case of statistically insufficient data samples, or spherically symmetric sources it is advantageous to measure the correlation functions versus a single-dimensional momentum variable. A natural choice may be the invariant momentum difference, equivalent to the magnitude of the three-momentum difference in the pair comoving (i.e. pair-center-of-mass) system (PCMS). Another possible choice is the magnitude of the three-momentum difference in the longitudinally comoving system (LCMS):
\begin{equation}
q \equiv q_{\textnormal{LCMS}}= \sqrt{(p_{1,x}-p_{2,x})^2+(p_{1,y}-p_{2,y})^2+q_{z,\textnormal{LCMS}}^2},
\end{equation}
where the coordinate system is set up such that $z$ is the direction of the beam, also sometimes called the longitudinal direction; and the transverse plane coordinates are $x$ and $y$, which can be chosen arbitrarily. The momentum difference in this direction can be expressed in the LCMS as:
\begin{equation}
q^2_{z,\textnormal{LCMS}} = 4\cdot \frac{(p_{z,1} \cdot E_2 - p_{z,2} \cdot E_1)^2}{(E_1 + E_2)^2 - (p_{z,1} + p_{z,2})^2},
\end{equation}
where $E_1$ and $E_2$ are the energies of the respective particles. The LCMS can be advantageous since hadron emission turns out to be approximately spherically symmetric in this frame at RHIC energies~\cite{PHENIX:2017ino} and at GSI, HADES as well~\cite{HADES:2019lek}. We note in passing that preliminary investigation of the full three-dimensional correlation function indicate that, indeed this is a natural variable for parametrizing Bose-Einstein correlations, also at SPS energies for Be+Be collisions.

Assuming a three-dimensional spherically symmetric L\'evy shaped source function and the core-halo model, the corresponding parametric form of the two-particle Bose-Einstein correlation function becomes
\begin{equation}
C_2(q) =1+\lambda\cdot e^{-|qR|^\alpha}.\label{e:fittingformulanocoul}
\end{equation}
Its three parameters $\lambda$, $R$, and $\alpha$ implicitly depend on the average transverse momentum \kt, or, alternatively, on the transverse mass \mt.

The shape parameter $\alpha$ carries information on the nature of the quark-hadron transition. Namely, lattice QCD calculations~\cite{Aoki:2006we,Bhattacharya:2014ara,Soltz:2015ula} and other theoretical expectations show two important regions of the baryochemical potential ($\mu_B$) axis of the phase diagram of strongly interacting matter. A phase transition of analytical or ``cross-over'' type is expected at low $\mu_B$ values and a first-order phase transition at high values of $\mu_B$. Therefore, a critical endpoint of the phase transition line is expected, where a second-order phase transition takes place. The mapping of the phase diagram of strongly interacting matter, and determining the position of the critical endpoint is one of the main goals of high-energy heavy-ion physics experiments, such as CBM~\cite{Agarwal:2023otg}, HADES~\cite{Stroth:2017blf}, PHENIX~\cite{PHENIX:1998vmi}, STAR~\cite{Yang:2021lfe}, and NA60+~\cite{NA60:2022sze}.

The L\'evy stability index $\alpha$ is related to the spatial critical exponent $\eta$~\cite{Csorgo:2008ayr}, since at the critical endpoint, fluctuations appear at all scales and spatial correlations will exhibit a power-law tail of the form $\sim r^{-1 -\eta}$, and L\'evy distributed sources also exhibit a power-law tail $\sim r^{-1 -\alpha}$ (in three dimensions).

Theoretical expectations suggest that the universality class of QCD to be the same as that of the 3D Ising model~\cite{Halasz:1998qr,Stephanov:1998dy}. The value of the exponent $\eta$ around the critical point in the 3D Ising model is 0{.}03631 $\pm$ 0.00003~\cite{El-Showk:2014dwa}, and with a random external field it is seen to be 0{.}50 $\pm$ 0.05~\cite{Rieger:1995aa}. This argument suggests that close to the critical endpoint (CEP) of the phase transition line of strongly interacting matter, $\alpha$ should also decrease to values near or even possibly below 0.5. While finite size effects and dynamics may modify this simple picture, measuring the L\'evy stability index $\alpha$ is still expected to provide a signature of the critical point of the phase diagram of strongly interacting matter.

\subsection{Final state Coulomb effect}\label{subsec:coulomb}
The final state phenomena, such as the electromagnetic interactions between charged hadrons, were neglected in the above considerations. Namely, the quantum-statistical correlation functions discussed so far were obtained with the plane-wave assumption for the wave function. In the following, these will be denoted by $C^0(q)$. 
Let’s suppose that the final-state electromagnetic interactions are included in the correlation function. Then the correlation function has to be calculated not via the interference of plane waves, but rather via the interference of solutions of the two-particle Schrödinger equation having a Coulomb-potential, describing the final state electromagnetic interactions. The ratio of these two correlation functions is called the Coulomb correction~\cite{Kincses:2019rug,Csanad:2019lkp}:

\begin{align}\label{e:coulcorr_equation}
K_{\textnormal{Coulomb}}(q) = \frac{C^{\textnormal{ Coul}}(q)}{C^0(q)}.
\end{align}

The numerator in Eq.~\eqref{e:coulcorr_equation} cannot be calculated analytically and is quite tedious to estimate numerically. To simplify experimental analysis, in Ref.~\cite{CMS:2017mdg}, an approximate formula was obtained and utilized subsequently for the case of Cauchy-shaped sources ($\alpha=1$). However, the Coulomb correction may also depend on the L\'evy stability index $\alpha$, hence a more precise treatment is required. To this end, a numerical calculation was performed in Refs.~\cite{Csanad:2019cns, Csanad:2019lkp} and the results were parameterized, then the dependence on \textit{R}, $\lambda$ and $\alpha$ has been parameterized as well. In this analysis, we utilize the results obtained in Refs.~\cite{Csanad:2019cns, Csanad:2019lkp} for estimating the Coulomb effect. 

In order to take into account the effect of the halo mentioned in Sec. \ref{s:corehalo}, the Bowler-Sinyukov method~\cite{Sinyukov:1998fc,Bowler:1991vx} is utilized. The halo part only contributes at very small values of relative momenta, and hence it does not affect the source radii of the core component~\cite{Maj:2009ue}. This justifies the mentioned Bowler-Sinyukov method, in which the fit ansatz function is as follows:

\begin{equation}
C_2(q) = N\cdot\left(1-\lambda+(1+e^{-|qR|^\alpha})\cdot \lambda\cdot K_{\textnormal{Coulomb}}(q)\right).\label{e:fittingformula}
\end{equation}

Here $K_{\textnormal{Coulomb}}(q)$ is the Coulomb correction, and a normalisation parameter $N$ was also introduced. 

In addition, another effect has to be taken care of, which is related to the fact that the Coulomb correction is calculated in the PCMS while the measurement is done in the LCMS. When measuring one-dimensional HBT correlations in the LCMS, the assumption is that the source is of spherical shape, meaning $R_{\textnormal{out}} = R_{\textnormal{side}} = R_{\textnormal{long}} = R \equiv R_{\textnormal{LCMS}}$.
However, the source is only spherical in the LCMS, hence an approximate one dimensional PCMS size parameter needs to be estimated.

This was done in Ref.~\cite{Kurgyis:2020vbz}, where an average PCMS radius of
\begin{equation}
\overline{R}_{\textnormal{PCMS}} = \sqrt{\frac{1-\frac{2}{3}\beta^2_{\textnormal{T}}}{1-\beta^2_{\textnormal{T}}}} \cdot R
\end{equation}
was obtained, where $\beta_{\textnormal{T}} = \frac{\kt}{\mt}$. Furthermore, the following fact has to be taken into account: the momentum difference in the Coulomb correction is expressed in the PCMS, as a function of $q_{\textnormal{PCMS}}=q_{\textnormal{inv}}$ (the invariant four-momentum difference equals the three-momentum difference in the PCMS). Since the reconstruction of $q_{\textnormal{inv}}$ for a given pair (knowing only $q$) is not possible, the measurement should be performed as a function of both $q$ as well as $q_{\textnormal{inv}}$. The estimation performed in Ref.~\cite{Kurgyis:2020vbz} showed that a simple, approximate relation of the two may be given as $q_{\textnormal{inv}} \approx \sqrt{1-\beta_{\textnormal{T}}^2/3}\cdot q$. Implementing both of the above mentioned effects results in the following formula for the Coulomb correction expressed in terms of $q$
and $R_{\textnormal{LCMS}}$, based on the 3D calculation in PCMS:
\begin{equation}\label{e:KCoulombfinal}
    K_{\textnormal{Coulomb}}\left(q,R\right) = 
    K_{\textnormal{Coulomb}}^{3D,\;PCMS}\left(\sqrt{1-\frac{\beta_{\textnormal{T}}^2}{3}}\cdot q, \sqrt{\frac{1-\frac{2}{3}\beta^2_{\textnormal{T}}}{1-\beta^2_{\textnormal{T}}}} \cdot R\right),
\end{equation}
where now the dependence on $R$ is indicated explicitly.
This modified Coulomb correction is then used in Eq.~\eqref{e:fittingformula}.
Note that $K_{\textnormal{Coulomb}}$ depends also on the Lévy-index $\alpha$, but the mentioned PCMS-LCMS transformation leaves this parameter unaffected, hence we suppressed this from the function arguments above. Furthermore, it should also be underlined that the effect of modifying the Coulomb correction based on the PCMS-LCMS difference discussed above is small, particularly negligible, compared to the listed systematic uncertainty sources of Section~\ref{s:systunc}.

\section{The \NASixtyOne detector}
\label{sec:NA61SHINE}
\begin{figure}[t!]
\centering
\includegraphics[width=.8\textwidth]{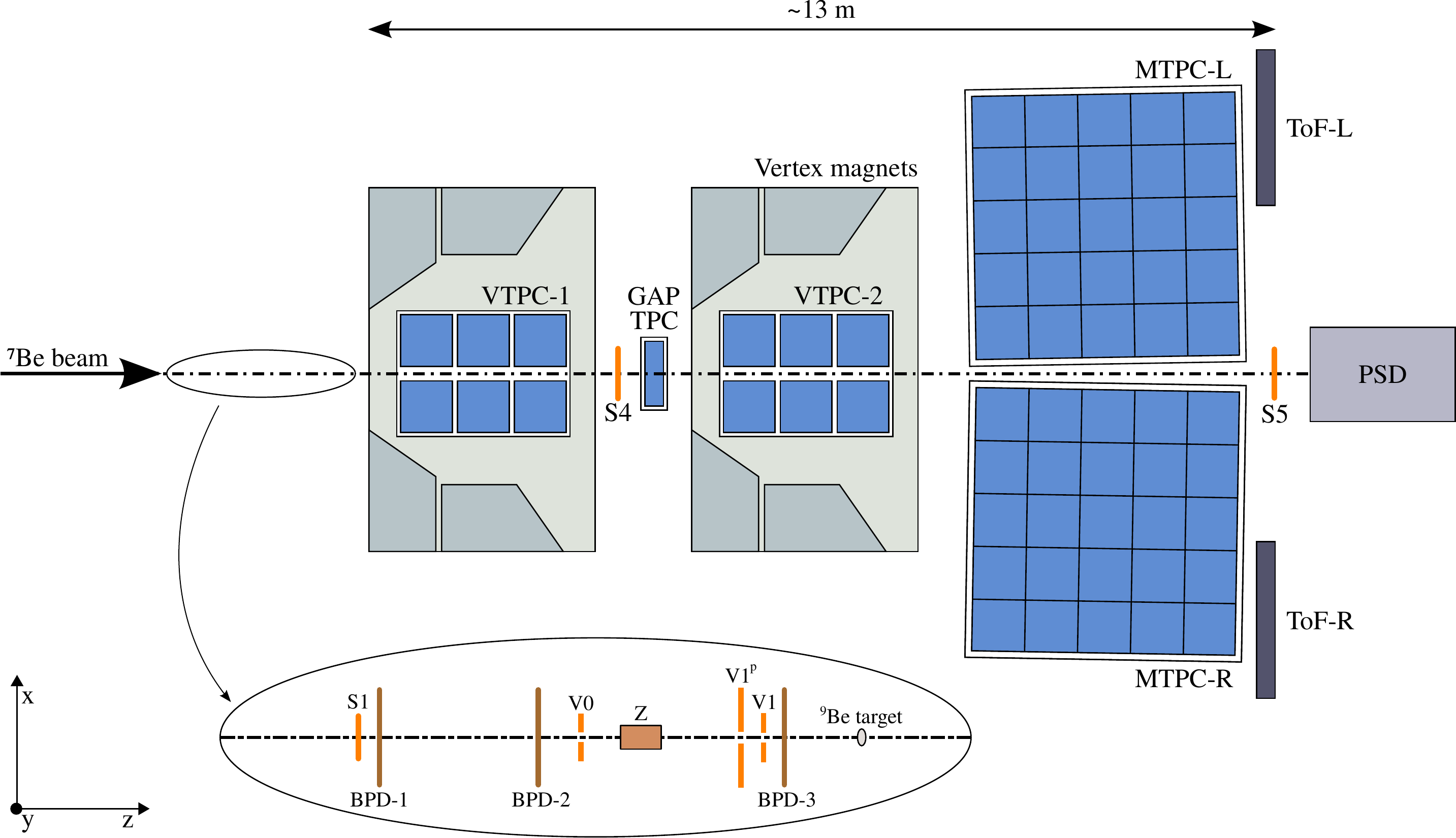}\\
\caption{Schematic of the \NASixtyOne detector setup, used during the Be+Be data taking.}
\label{fig:na61_bebe}
\end{figure}

The \NASixtyOne fixed-target experiment uses a large acceptance hadron spectrometer located in the North Area H2 beam line of the CERN Super Proton Synchrotron accelerator~\cite{Abgrall:2014xwa}. The main goals of the experiment include the investigation of the phase diagram of strongly interacting matter. A schematic of the layout employed during the Be+Be data taking is shown in Fig.~\ref{fig:na61_bebe}.
\subsection{Detectors}\label{s:detectors}
The key components of the experiment for the detection of particles produced in the collisions are the five large-volume Time Projection Chambers (TPCs) for tracking. The two most upstream ones are the Vertex TPCs (VTPCs), residing in the two superconducting bending magnets. The magnets have 9 T$\cdot$m maximum combined bending power. Downstream of the VTPCs, the two Main TPCs (MTPCs) are located symmetrically to the beam line in order to extend the tracking lever arm and to perform particle identification by measuring their ionisation energy loss in the TPC gas.  One smaller TPC is located in the gap between VTPCs, and is called Gap-TPC (GTPC; denoted GAP TPC in Fig.~\ref{fig:na61_bebe}). The VTPCs and GTPC are operated with an Ar(90):CO$_2$(10) gas mixture and the MTPCs with an Ar(95):CO$_2$(5) mixture. The further downstream Time-of-Flight (ToF) detectors are not used in the present analysis.

The Projectile Spectator Detector (PSD) at the end of the setup is a segmented forward hadron calorimeter, centered on the nominal deflected beam trajectory. It measures the energy contained in the projectile spectators which is used for event centrality characterization. Central collisions are selected by lower values of this very forward energy. Although for smaller systems, such as Be+Be collisions, the very forward energy is not expected to be tightly correlated with the actual impact parameter of the collision, the terms central and centrality are still adopted following the convention widely used in heavy-ion physics.

The beam line instrumentation is schematically shown in the inset of Fig.~\ref{fig:na61_bebe}.
A set of scintillation counters as well as beam position detectors (BPDs)~\cite{Abgrall:2014xwa} upstream of the target
provide timing reference, selection, identification and precise measurement of the position
and direction of individual beam particles.

\subsection{Triggers} 
The schematic of the placement of the beam and trigger detectors is shown in the inset of Fig.~\ref{fig:na61_bebe}.
These consist of a scintillation counter (S1) recording the presence of the beam particle, a set of veto scintillation counters with hole (V0, V1, V1$^p$) used for rejecting beam halo particles, and a threshold Cherenkov charge detector (Z). Trigger signals indicating the passage of valid beam particles are defined by the coincidence T1 = $\text{S1} \cdot \overline{\text{V0}} \cdot \overline{\text{V1}} \cdot \overline{\text{V1$^p$}} \cdot \text{Z(Be)}$ for high momentum data taking.

\textit{Central} collisions were selected through the analysis of the signal from the 16 central modules of the PSD \cite{EmilThesis}. The low-energy part of the deposited energy spectrum was selected to contain 20\% of the most central collisions. 
The interaction trigger condition was thus T2 = T1$\cdot\overline {\text{PSD}}$ for the higher energies. 

The data consists of $\approx 2{.}828\cdot 10^6$ events before event and track selection.

\subsection{The $^7$Be beam and $^9$Be target}
The \NASixtyOne beam line is designed to handle primary as well as secondary beams. 
The beam instrumentation was optimized accordingly. In the Be+Be runs, a secondary beam was used, fragmented from a primary Pb beam from the SPS accelerator. A threshold Cherenkov charge tagging detector, called the Z detector, was used in order to identify and select the $Z=4$ fragment nuclei. In order to have a low material budget for the Z detector, a thin quartz wafer Cherenkov radiator was used.
Additionally, the amplitudes of the signals measured in the three Beam Position Detectors (BPDs, see Fig.~\ref{fig:na61_bebe}) were used to improve the Z resolution. A detailed description of the technique for the identification of $^7$Be fragments is given in Ref.~\cite{Gazdzicki:1322135}. 

The target was a 12 mm thick plate of $^9$Be placed approximately 80.0 cm upstream of VTPC-1. The total mass concentrations of impurities in the target were measured at 0.287\% ~\cite{Banas:2018sak}. No correction was applied for this negligible contamination.

%***********************************************************************************
\section{Analysis procedure}
\label{sec:Analysis}
Femtoscopic correlations are studied in this paper for pions reconstructed as originating from the primary interaction in the 20\% most central $^{7}$Be+$^{9}$Be collisions selected by the total energy emitted into the forward direction covered by the PSD detector. In the following, we describe the event, track and pair selection procedure, and all the steps required to obtain the measured source parameters. 

\subsection{Event selection}
The events considered for analysis had to satisfy the following conditions:
\begin{enumerate}[(i)]

    \item there are no off-time beam particles detected within a time window of $4.5~\mu$s around the particle triggering the event (this is the time needed to have good position resolution of the main vertex at all times),
    %\footnote{Particles coming within the drift time of the chambers are considered off-time as the detector system is not emptied out. This 4.5 $\mu$s window is the time needed to having covered enough \textit{y} coordinate distance (drift direction) compared to the position resolution of main vertex reconstruction.}
    \item the event has a well-fitted main interaction vertex,
    \item the maximal distance between the main vertex $z$ position and the centre of the beryllium target is between $\pm$ 5~cm (vertex $z$),
    \item the 0--20\% most central collisions, based on PSD energy measurement, are accepted.
\end{enumerate}

\subsection{Track selection}\label{s:trackselect}
The tracks selected for the analysis had to satisfy the following conditions:

\begin{enumerate}[(i)]
    \item the fit of the particle track converged,
    \item the distance between the track extrapolated to the interaction plane and the interaction point (impact parameter) should be smaller or equal to 4~cm in the horizontal (bending -- $|B_{x}|$) plane and 2~cm in the vertical (drift -- $|B_{y}|$) plane\footnote{Track impact point resolution depends on track multiplicity in the event.},

    \item the total number of reconstructed points in all TPCs on the track should be at least 30 (nPoint) and, at the same time, the sum of the number of reconstructed points in VTPC-1 and VTPC-2 should be at least 15 (VTPC points) or the number of reconstructed points in the GTPC should be at least 5 (GTPC points),
    
    \item the ratio of the total number of reconstructed points on the track to the potential number of points should be between 0.5 and 1.0\footnote{Due to uncertainty of the momentum fitting and the fitted interaction point, the nPointRatio values may exceed 1. Hence, the upper limit for the ratio was set to 1.2 when estimating systematic uncertainties.} (nPointRatio),
    
    \item identified particle's rapidity is in the interval $\pm2$ around midrapidity.
\end{enumerate}

\subsection{Particle identification}

\begin{figure}
\begin{center}
\includegraphics[width=0.49\textwidth]{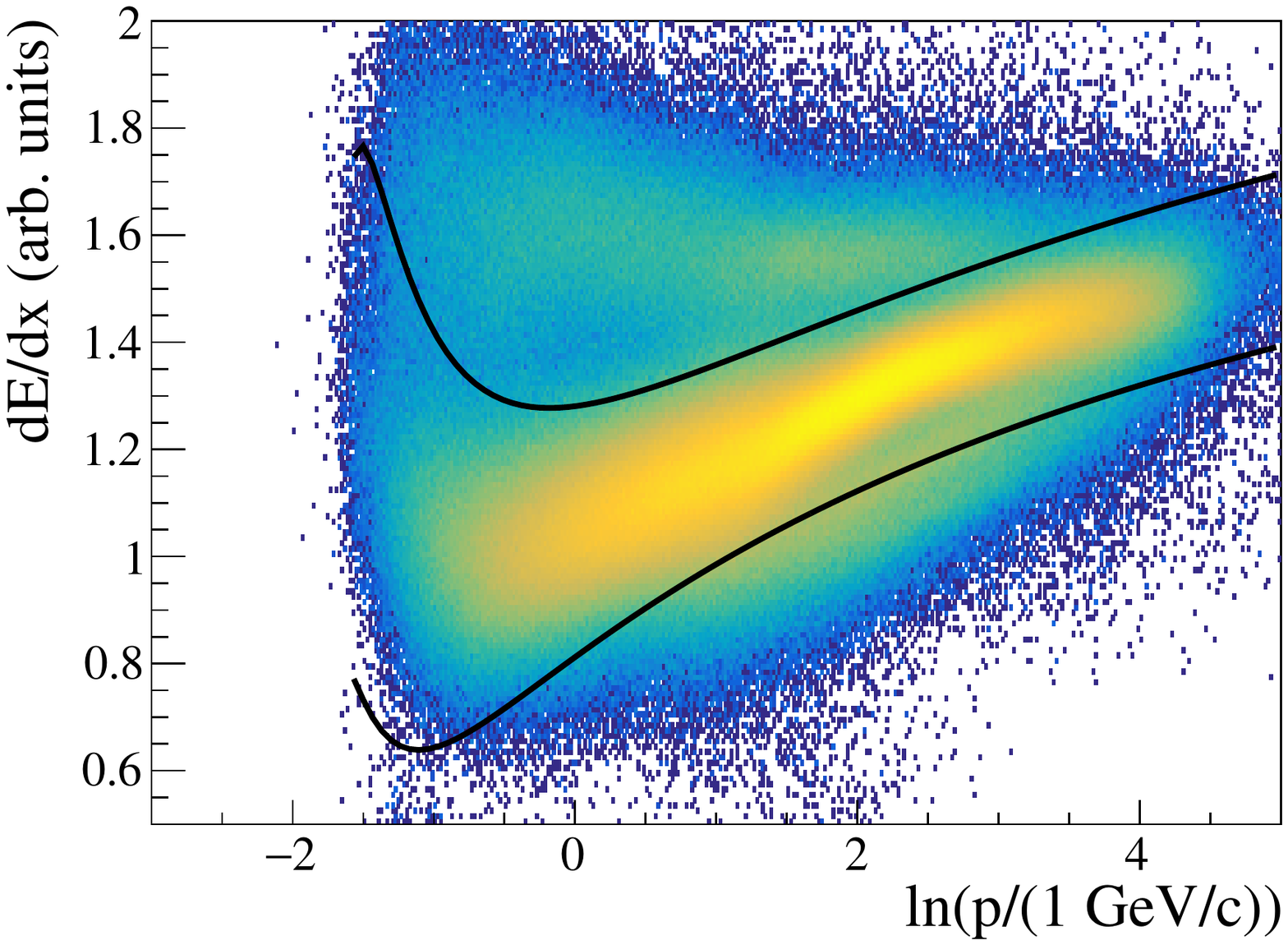}
\includegraphics[width=0.49\textwidth]{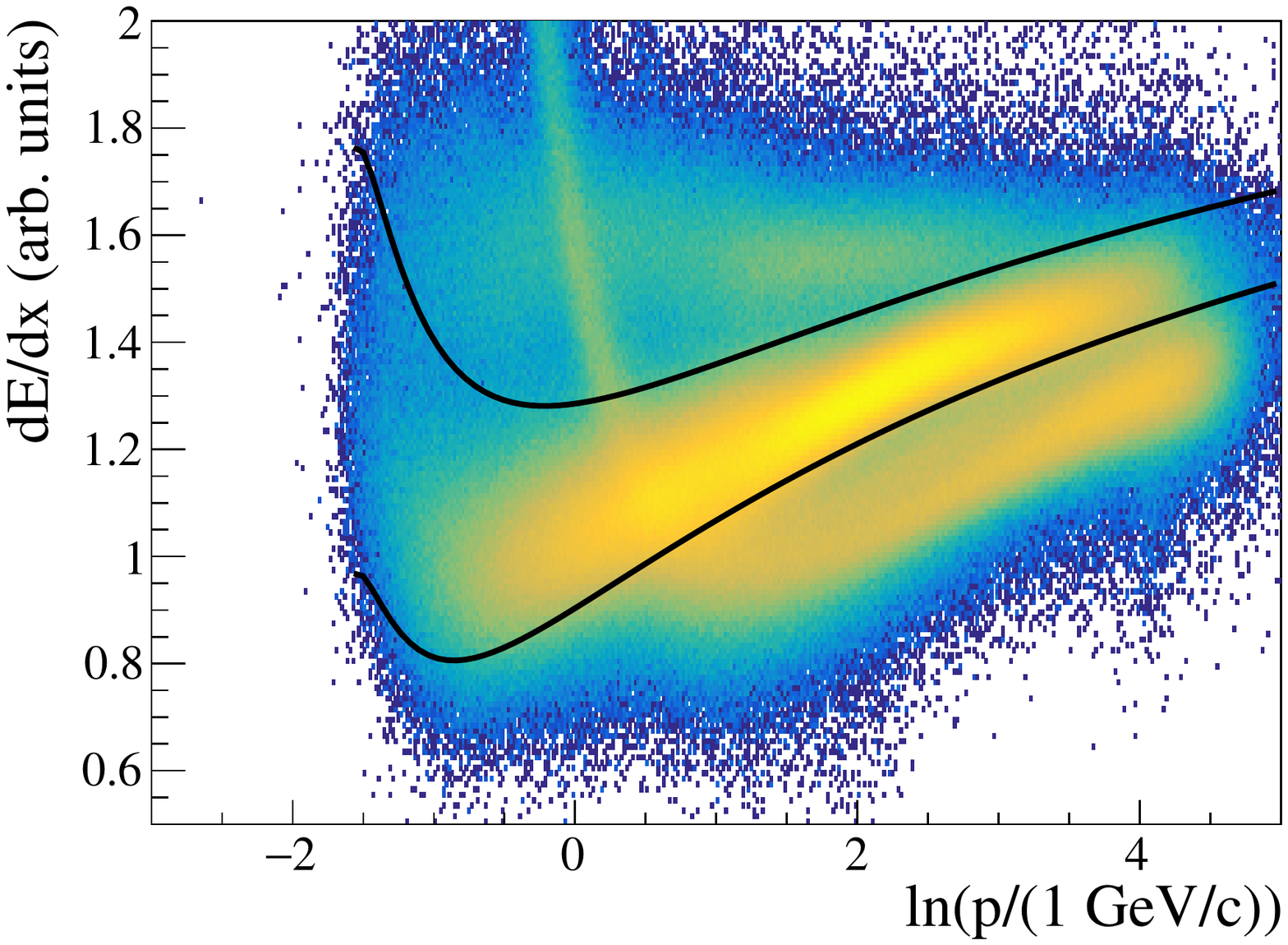}
\end{center}
\caption{The \dedx measurement for negatively charged (left panel) and positively charged (right panel) particles versus natural logarithm of momentum $p$ (in laboratory frame). The two lines represent the interpolated selection boundaries based on Gaussian fits to \dedx distributions at several momenta, and are 3 standard deviations from the pion mean for negatively charged particles, while for positively charged particles, the lower line is 1{.}5 standard deviations distance from the mean, to remove the more significant kaon and proton contributions present in this case.}
\label{fig:pidplots_BeBe}
\end{figure}

Particle identification (PID) for pions was performed using \dedx cuts, shown in Fig.~\ref{fig:pidplots_BeBe}. Tracks within the two lines are considered pions. The fine-tuning of the \dedx cut parameters was performed as follows.
\begin{enumerate}[(i)]
    \item a reasonable interval in $\ln(p/1 (\textnormal{GeV}/c)$ was selected where the pion contribution dominates (from -1.6 to 4, ~0.2 to 55 \GeVc),
    \item the data was binned in $\ln(p)$ into 80 slices,
    \item in each bin, a Gaussian was fitted to the \dedx data, in order to establish the most probable value of the pion \dedx peak,
    \item the standard deviation ($\sigma$) obtained from these Gaussian fits was used to determine the pion \dedx response width (found to be between 0.05 and 0.18, depending on momentum), which was the basis of the pion \dedx selection as shown in Fig.~\ref{fig:pidplots_BeBe} (PID cut).
\end{enumerate}

\subsection{Pair selection and event mixing}
Due to the possible imperfections of the detector and  of the tracking algorithm, hits created by a single particle may be reconstructed as two tracks. This is called track splitting and leads to a track pair with small momentum difference (< 20 \MeVc). Furthermore, the hits of two close particles may be reconstructed as a single track: this is called track merging. In a correlation analysis, it is important to minimize the effect of these track reconstruction problems. Track splitting is already largely removed by the track selection cuts (in particular, (iv) of Sec.~\ref{s:trackselect}). The contribution from track merging was estimated by Monte Carlo (MC) simulations using EPOS simulation~\cite{Pierog:2009zt} and GEANT3 for particle propagation~\cite{Geant3} and reconstruction, and an appropriate lower limit in the momentum difference was defined, as described below. 

The basic quantity of correlation measurements is the pair distribution. From pairs of pions created in the same event, one obtains the so-called actual pair distribution $A(q)$. Such distributions were measured in several intervals of average pair transverse momentum \kt or pair transverse mass \mt. This pair distribution is influenced by single-particle momentum distributions, kinematic acceptance of the detector, phase-space effects, and other phenomena not connected to quantum-statistics or final state interactions. These can be removed by constructing a combinatorial background pair distribution $B(q)$, measured in the same \kt or \mt intervals as the $A(q)$ distribution.  Calculating this background distribution starts with the event mixing procedure, where an artificial, mixed event is created from particles originating from different events. Subsequently, pairs formed within the mixed event are used to create the background distribution $B(q)$. By construction, this method ensures that no two particles are selected from the same background event, creating an uncorrelated pool of events.

\begin{figure}
\includegraphics[width=0.48\textwidth]{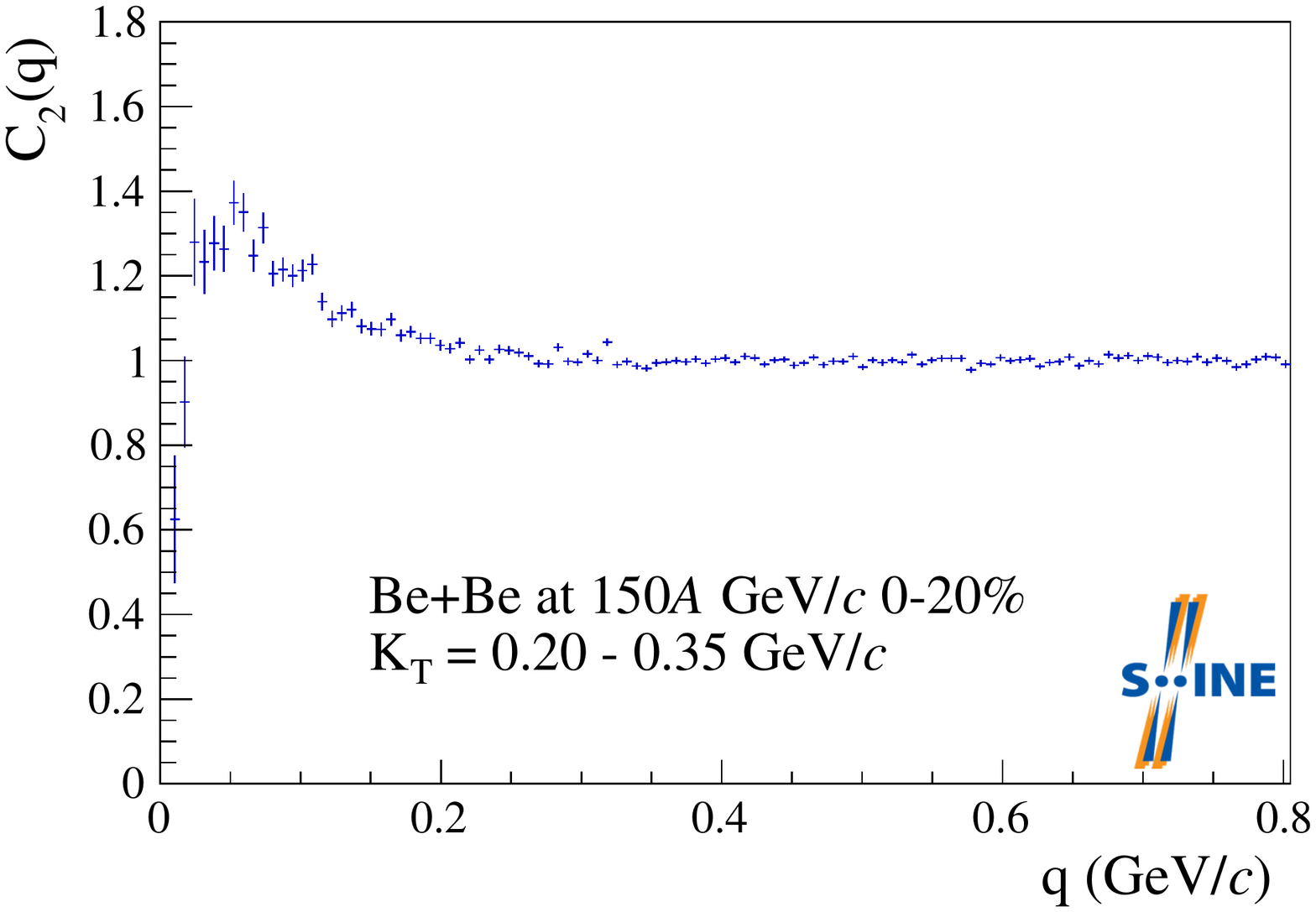}
\includegraphics[width=0.48\textwidth]{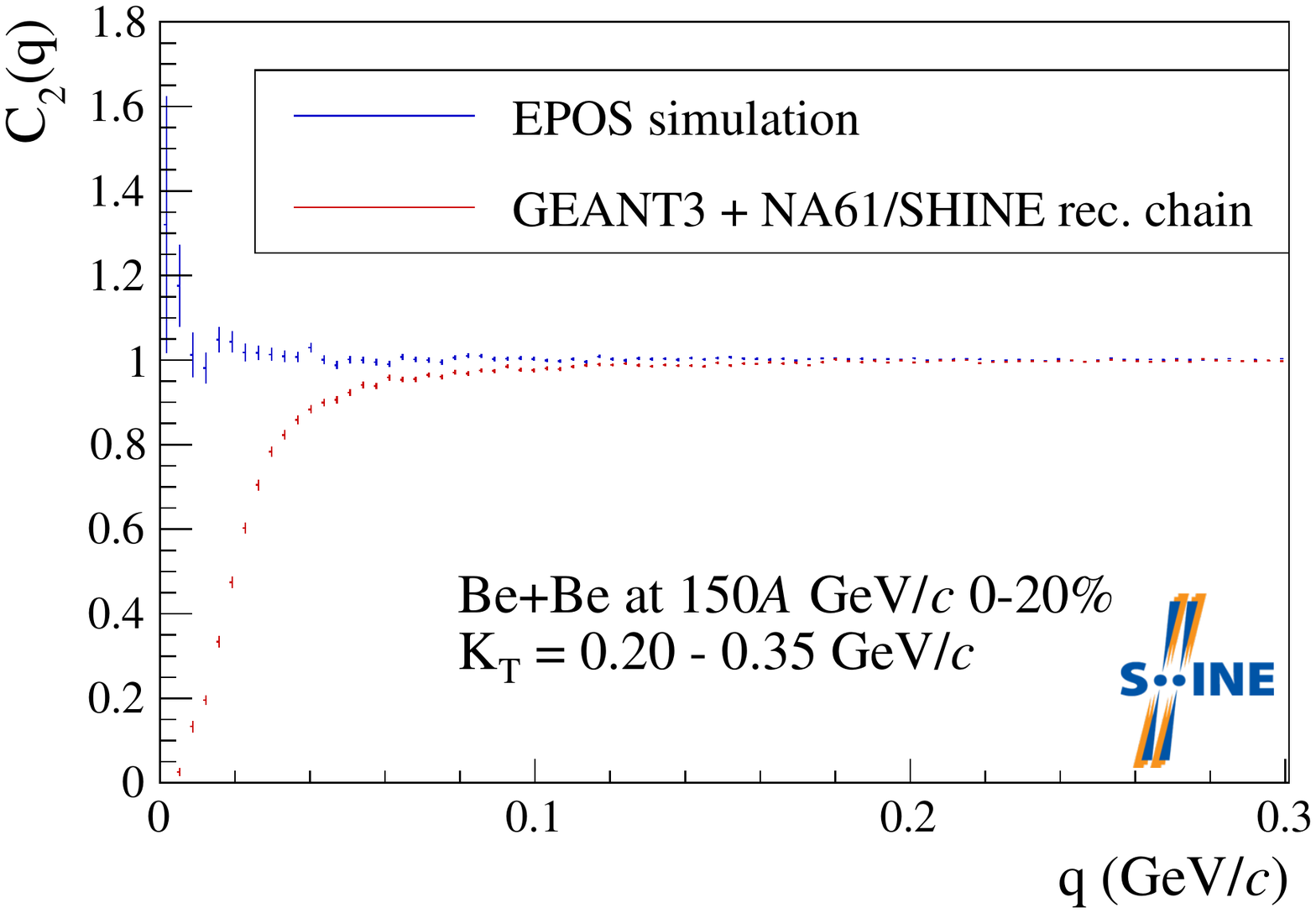}
\caption{The ratio of $A(q)$ and $B(q)$ as a function of $q$. Left: measured NA61/SHINE data. Right: EPOS simulation and EPOS reconstructed data (GEANT3 + NA61/SHINE rec. chain).}
\label{fig:C_data_epos}
\end{figure}

The obtained background distribution $B(q)$ exhibits all the previously mentioned non-quantum-statistical effects (acceptance, momentum distribution, phase-space, etc.), hence dividing $A(q)$ by $B(q)$ leaves us with a ratio which exhibits quantum-statistical and final-state interaction effects as well as the effect of reconstruction inefficiencies (and, in addition, momentum conservation, which is not relevant in the range of the investigated $q$-range). Thus, the measured correlation function is defined as

\begin{equation}
C_2(q) = \frac{A(q)}{B(q)}\cdot \frac{\int_{q_1}^{q_2}B(q)dq}{\int_{q_1}^{q_2}A(q)dq},\label{e:corrfuncAB}
\end{equation}
where [$q_1,q_2$] is a large-$q$ range where quantum-statistical effects no longer affect the correlation function. The integrals in Eq.~\eqref{e:corrfuncAB} provide the normalization of the correlation function to unity at high relative momentum.
An example for $C_2(q)$ is shown in Fig.~\ref{fig:C_data_epos} for both data and EPOS simulation. It is readily apparent that at low $q$ values Bose-Einstein correlation and Coulomb repulsion effects determine $C_2(q)$ data points. These effects are not present in the simulations, hence the simulated $C_2(q)$ values are approximately constant. The reconstructed $C_2(q)$, however, suffers from track merging effects (where the two tracks forming the pairs are close spatially), strongly suppressing $C_2(q)$ at low $q$ values. Hence the deviation of the simulated and reconstructed correlation function provides a good estimate of the range where inefficiencies are important. This allows to determine the range in $q$ over which fits can be considered reliable. The fit range is then selected for each $K_{\textnormal{T}}$ bin, e.g. for $K_{\textnormal{T}} = 0.20\--0.35$ \GeVc the interval where fit is considered good is $q=0.049 \-- 0.8$ \GeVc. This method then ensures that track merging is not present in the fitting interval.

\subsection{Estimation of source shape parameters via fitting}\label{ss:fitting}
\begin{figure}
\centering
\includegraphics[width=.85\textwidth]{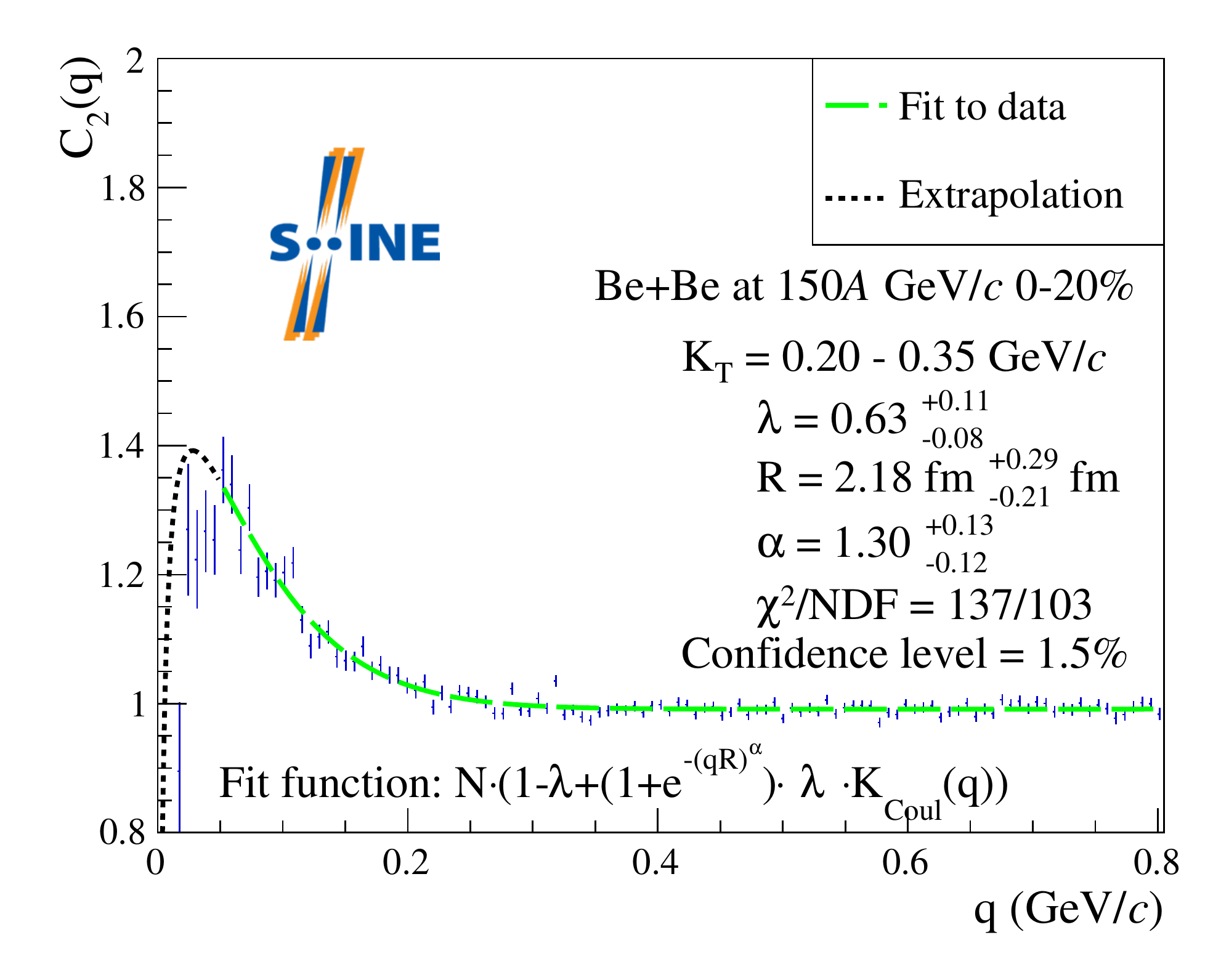}
\caption{Example fit with Bose-Einstein correlation function at $\kt = 0.20\--0.35$~\GeVc for the sum $\left(\pi^++\pi^+\right) + \left(\pi^-+\pi^-\right)$. Blue points with error bars represent the data, the green dash-dotted line shows the fitted function with Coulomb correction given by Eq.~\eqref{e:fittingformula} within the range of 0.049~\GeVc to 0.8~\GeVc, and the black dotted line indicates the extrapolated function outside of the fit range.}
\label{fig:examplefit}
\end{figure}

The measured correlations were fitted with the formula described in Eq.~\eqref{e:fittingformula}. Due to the often modest number of entries in the signal and background distributions~\cite{zajc1984two}, Poisson maximum-likelihood fitting was used~\cite{Baker:1983tu}. The corresponding penalty function ($\chi^2_{\lambda,p}$) to minimise is

\begin{equation}
\chi^2_{\lambda,p} = 2 \Big[\sum_i \Big( y_i - n_i \Big) + \Big(\sum_{\mathclap{\substack{i\\c_i \neq 0}}} n_i\cdot \ln(n_i/y_i)\Big)\Big],
\end{equation}

where $\lambda$ and $p$ denote the fact that we are using a likelihood $\chi^2$ for Poisson distributed histograms, $c_i$ references the number of counts, $n_i$ is the number of entries in the i$^{th}$ histograming bin obtained from the data, and $y_i$ is its corresponding parametric model value to be fitted to the data. Goodness-of-fit was determined using regular $\chi^2$ methods in two ranges: the full range and the Bose-Einstein peak range. Fits were done both for positively and negatively charged pion pairs, as well as their combinations, in four \mt intervals. A fit was accepted if the algorithm converged, the covariance matrix was positive definite, and the confidence value corresponding to the $\chi^2\textnormal{ and NDF}$ was larger than 0.1\%. An example fit is shown in Fig.~\ref{fig:examplefit}. To estimate the statistical uncertainties of the fit parameters, the Minos method was utilised ~\cite{STATMETHODEXPPH01} (also called as likelihood based confidence intervals) which by its nature, yields asymmetric statistical errors.

\subsection{Systematic uncertainties}\label{s:systunc}

In the analysis, one has to consider that the parameters obtained from fits depend on several experimental choices and cuts, such as the PID cut, the width of bins, or the fitting range. These dependencies are the dominating contributors to the systematic uncertainties. In order to estimate these, the fits were performed with the loose and tight event and track selection criteria, and also with slightly varied fit intervals. The standard set of cut values together with the alternative values for systematic error estimation are shown in Table~\ref{tab:standard_loose_tight}. The systematic uncertainty calculation was performed for positively and negatively, like-sign charged pairs summed together.

\begin{table}[h!]
\centering
\begin{tabular}{|l|l|lll|}
\hline
\textbf{n} & \textbf{Source} & \multicolumn{1}{l|}{\textbf{standard}} & \multicolumn{1}{l|}{\textbf{tight}} & \textbf{loose} \\ \hline
0          & nPoint          & \multicolumn{1}{l|}{$\geq 30$}         & \multicolumn{1}{l|}{$\geq 40$}      & $\geq 10$      \\ \hline
1          & nPointRatio     & \multicolumn{1}{l|}{$0.5\--1.0$}         & \multicolumn{1}{l|}{$0.7\--1.0$}      & $0.4\--1.2$      \\ \hline
2          & VTPC points     & \multicolumn{1}{l|}{$\geq 15$}         & \multicolumn{1}{l|}{$\geq 30$}      & $> 10$         \\ \hline
3          & GTPC points     & \multicolumn{1}{l|}{$\geq 5$}          & \multicolumn{1}{l|}{$\geq 5$}       & $> 6$          \\ \hline
4 &
  \begin{tabular}[c]{@{}l@{}}$|B_x|$\\ $|B_y|$\end{tabular} &
  \multicolumn{1}{l|}{\begin{tabular}[c]{@{}l@{}} $\leq 4$ cm\\ $\leq 2$ cm\end{tabular}} &
  \multicolumn{1}{l|}{\begin{tabular}[c]{@{}l@{}} $\leq 0.8$ cm\\$\leq 0.8$ cm\end{tabular}} &
  \begin{tabular}[c]{@{}l@{}} $\leq 6$ cm\\ $\leq 5$ cm\end{tabular} \\ \hline
5          & $q$ bin width   & \multicolumn{1}{l|}{7 MeV/$c$}         & \multicolumn{1}{l|}{3.5 MeV/$c$}    & 3.5 MeV/$c$    \\ \hline
6          & Fit range       & \multicolumn{3}{l|}{\kt dependent, lower value is from MC}                                  \\ \hline
7 &
  PID cut &
  \multicolumn{1}{l|}{\begin{tabular}[c]{@{}l@{}}$\pi^+$: +3$\sigma$ and -1.5$\sigma$\\ $\pi^-$: +3$\sigma$ and -3$\sigma$\end{tabular}} &
  \multicolumn{1}{l|}{\begin{tabular}[c]{@{}l@{}}$\pi^+$: +2$\sigma$ and -1$\sigma$,\\ $\pi^-$: +2$\sigma$ and -2$\sigma$\end{tabular}} &
  \begin{tabular}[c]{@{}l@{}}$\pi^+$: +4$\sigma$ and -2$\sigma$\\ $\pi^-$: +4$\sigma$ and -4$\sigma$\end{tabular} \\ \hline
8          & vertex $z$ (cm) & \multicolumn{1}{l|}{-585-- -575}       & \multicolumn{1}{l|}{-585-- -575}               & -595 -- -565   \\ \hline
\end{tabular}
  \caption{The standard cuts used to obtain the final results, as well as the loose and tight cuts applied for estimation of systematic uncertainties.}
  \label{tab:standard_loose_tight}
\end{table}

The combined systematic uncertainties were obtained as follows. Let $P$ denote the fit parameter vector ($\alpha,\lambda,R$). Denote by $P_n^j(i)$ the corresponding estimated parameter vector obtained for the $i$-th \mt bin ($i=0,\dots,3$), with the $n$-th selection criterion ($n=0,\dots,8$) listed in Table~\ref{tab:standard_loose_tight} set to the $j$-th setting ($j=0,1,2$ meaning the standard, tight and loose values). The downward ($\delta P^-$) and upward ($\delta P^+$) systematic uncertainty of the parameter vector $P$ was estimated as follows:

\begin{align}
\delta P^+(i) = \sqrt{\sum_n \frac{1}{N^{j+}_n} \sum_{j \epsilon J_n^+} (P_n^j(i) - P^0(i))^2},\\
\delta P^-(i) = \sqrt{\sum_n \frac{1}{N^{j-}_n} \sum_{j \epsilon J_n^-} (P_n^j(i) - P^0(i))^2},
\end{align}

where $P^0(i)$ is the parameter vector in $i$-th \mt bin with standard cut ($j=0$),  $J_n^+$ and $J_n^-$ are the array of $j$ values with which $P^j_n(i) > P^0(i)\textnormal{, and } P^j_n(i) < P^0(i)$ occurs respectively, and $N^{j+}_n \textnormal{ and }N^{j-}_n$ denote their corresponding multiplicity.

%***********************************************************************************

\section{Results}
\label{sec:Results}

The three physical parameters ($\alpha,\;\lambda\; \textnormal{and}\; R$) were measured in four bins of pair transverse momentum \kt or pair transverse mass \mt. The parameters were obtained via fitting a parametric L\'evy ansatz on the source via the formula Eq. \eqref{e:fittingformula} to the measured correlation functions.

\begin{figure}[h!]
     \centering
     \includegraphics[width=.9\textwidth]{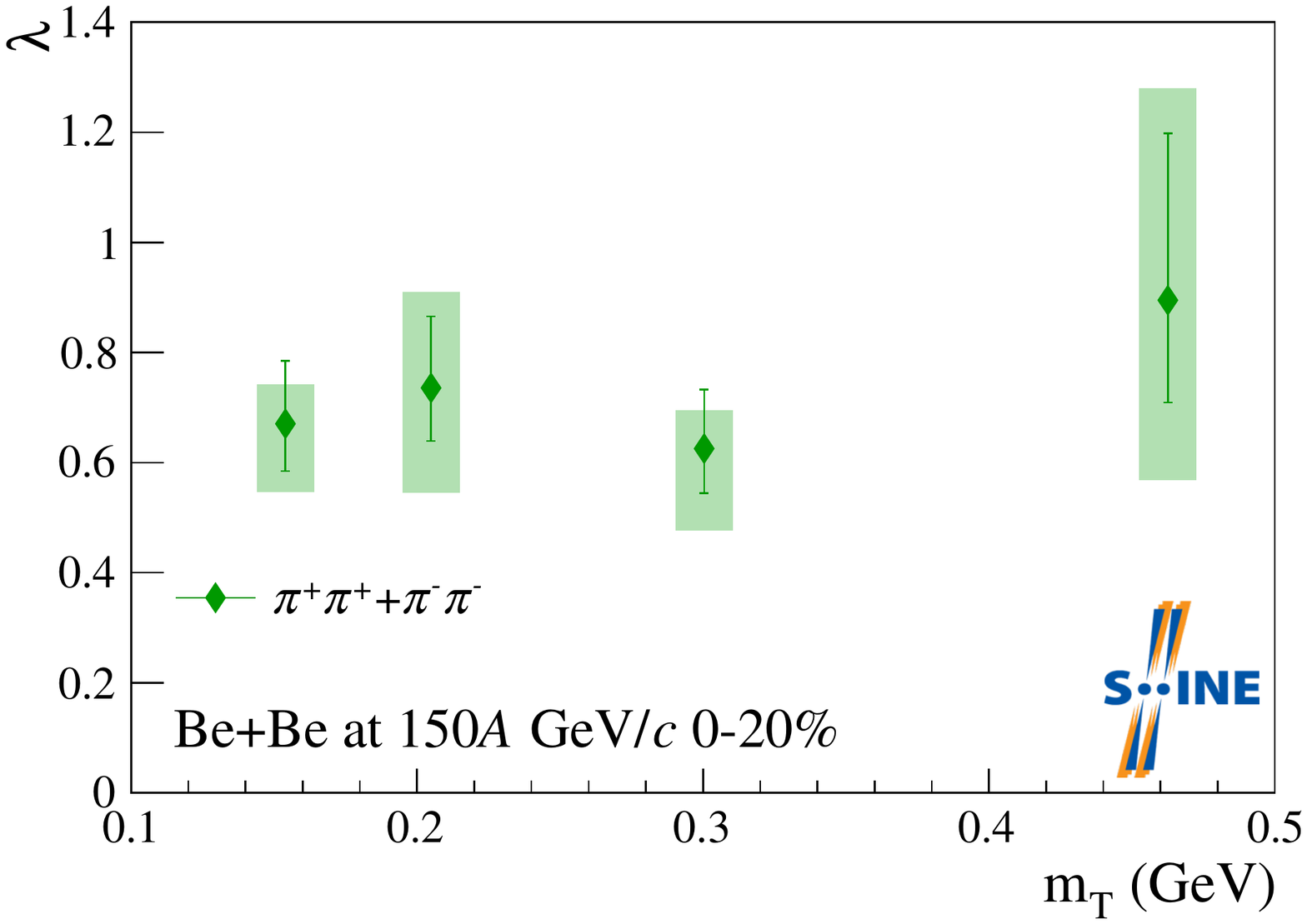}
     \caption{The intercept parameter $\lambda$, for 0--20\% central Be+Be at 150\AGeVc, as a function of \mt. Boxes denote systematic uncertainties, bars represent statistical ones.}
     \label{fig:lambda}
\end{figure}

The transverse mass dependence of the intercept parameter $\lambda$ is shown in Fig.~\ref{fig:lambda}. One may observe, within uncertainties, $\lambda(\mt)\approx$ const.\ in the available \mt range. An interesting phenomenon becomes apparent when compared to measurements at RHIC energy Au+Au collisions~\cite{PHENIX:2017ino,Vertesi:2009wf, STAR:2009fks} and at SPS energy S+Pb and Pb+Pb collisions~\cite{Beker:1994qv,NA49:2007fqa}. At the SPS energies, there is no visible decrease of $\lambda$ at lower \mt values, but at RHIC energies, a ``hole'' appears at \mt values around 2--300 \MeV. This ``hole'' was interpreted in Refs.~\cite{PHENIX:2017ino, Vance:1998wd} to be a sign of in-medium mass modification of the $\eta'$ meson. The \NASixtyOne results do not indicate the presence of a low-\mt hole. Furthermore, it is important to note that the obtained values for $\lambda$ are smaller than unity, which in the framework of the core-halo model may indicate that a significant fraction of pions are the decay products of long-lived resonances.

\begin{figure}[h!]
     \centering
\includegraphics[width=.9\textwidth]{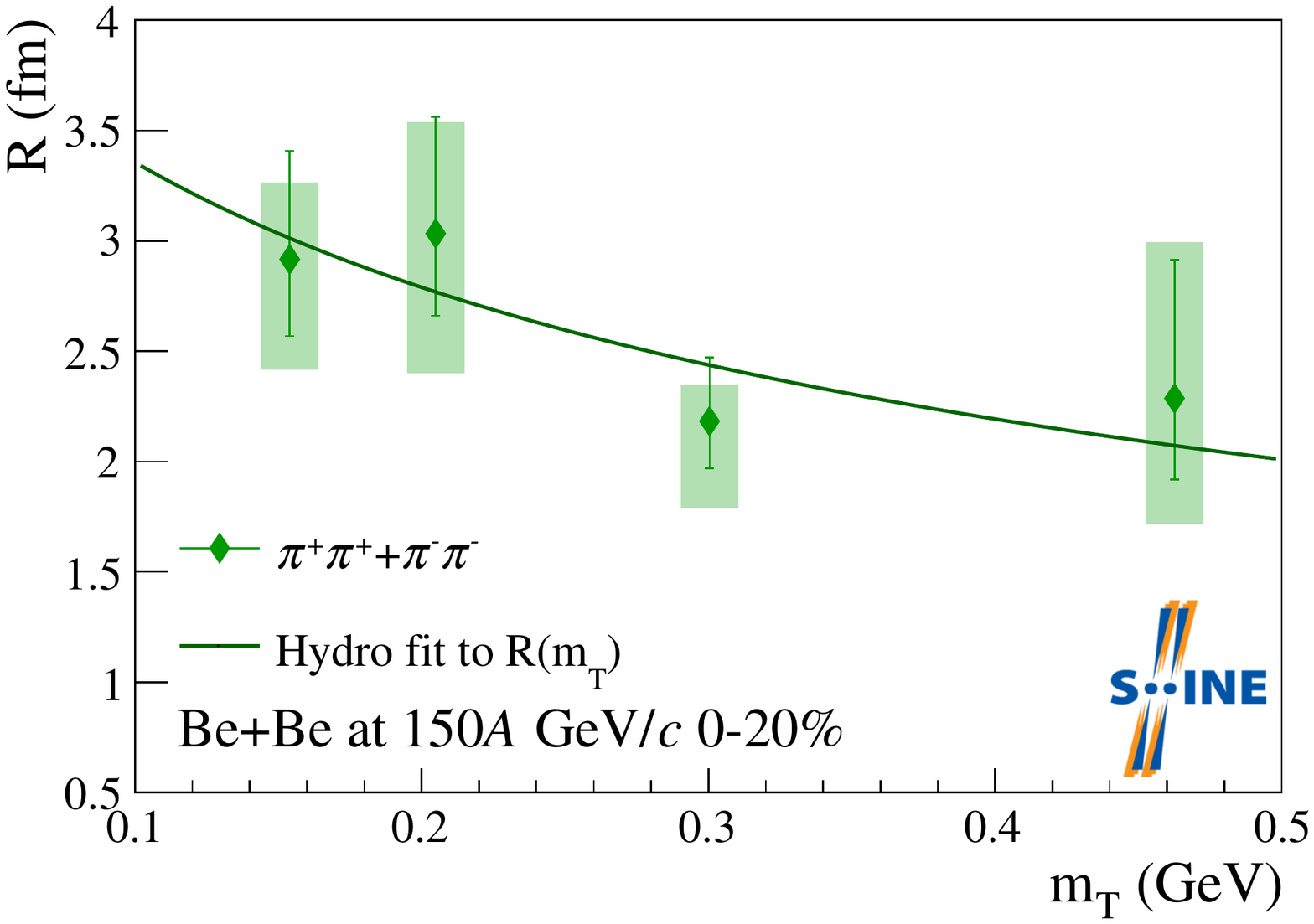}
     \caption{The radial scale parameter $R$, for 0--20\% central Be+Be at 150\AGeVc, as a function of \mt. The fit to $R(\mt)$ with Eq.~\eqref{e:r_theor} is shown with a solid line. Boxes denote systematic uncertainties, bars represent statistical ones.}
     \label{fig:R}
\end{figure}

The measured values of the radial scale parameter $R$ of the L\'evy shaped source function, determining the homogeneity length of the pion emitting source in the LCMS, are shown in Fig.~\ref{fig:R} as a function of \mt. Interestingly, the resulting $R$ parameter values are similar to those measured in \pp collisions at the CMS~\cite{CMS:2017mdg,CMS:2019fur}. We also observe a slight decrease of $R$ with increasing \mt. This can be explained by the presence of radial flow, based on simple hydrodynamical models~\cite{Csorgo:1995bi,Csanad:2009wc} where one obtains a $1/R^2 \propto \mt$ type of transverse mass dependence:
\begin{equation}\label{e:r_theor}
R = \frac{A}{\sqrt{1 + \mt / B}}.
\end{equation}
This function was fitted to the $R$ values measured in each $m_{\textnormal{T}}$ bin, as shown in Fig.~\ref{fig:R}, resulting in a good fit quality ($\chi^2/\textnormal{NDF}=1.7/2$, corresponding to a confidence level of 44\%). The obtained fit parameters are: $A = 4.5 \pm 2.9$ (stat.) $\textnormal{fm}$ and $B = 0.12 \pm 0.23$ (stat.) \GeV, comparable to those measured in \pp collisions at CMS~\cite{CMS:2019fur} (although the large uncertainties prevent a quantitative comparison).

\begin{figure}[h!]
     \centering
     \includegraphics[width=.9\textwidth]{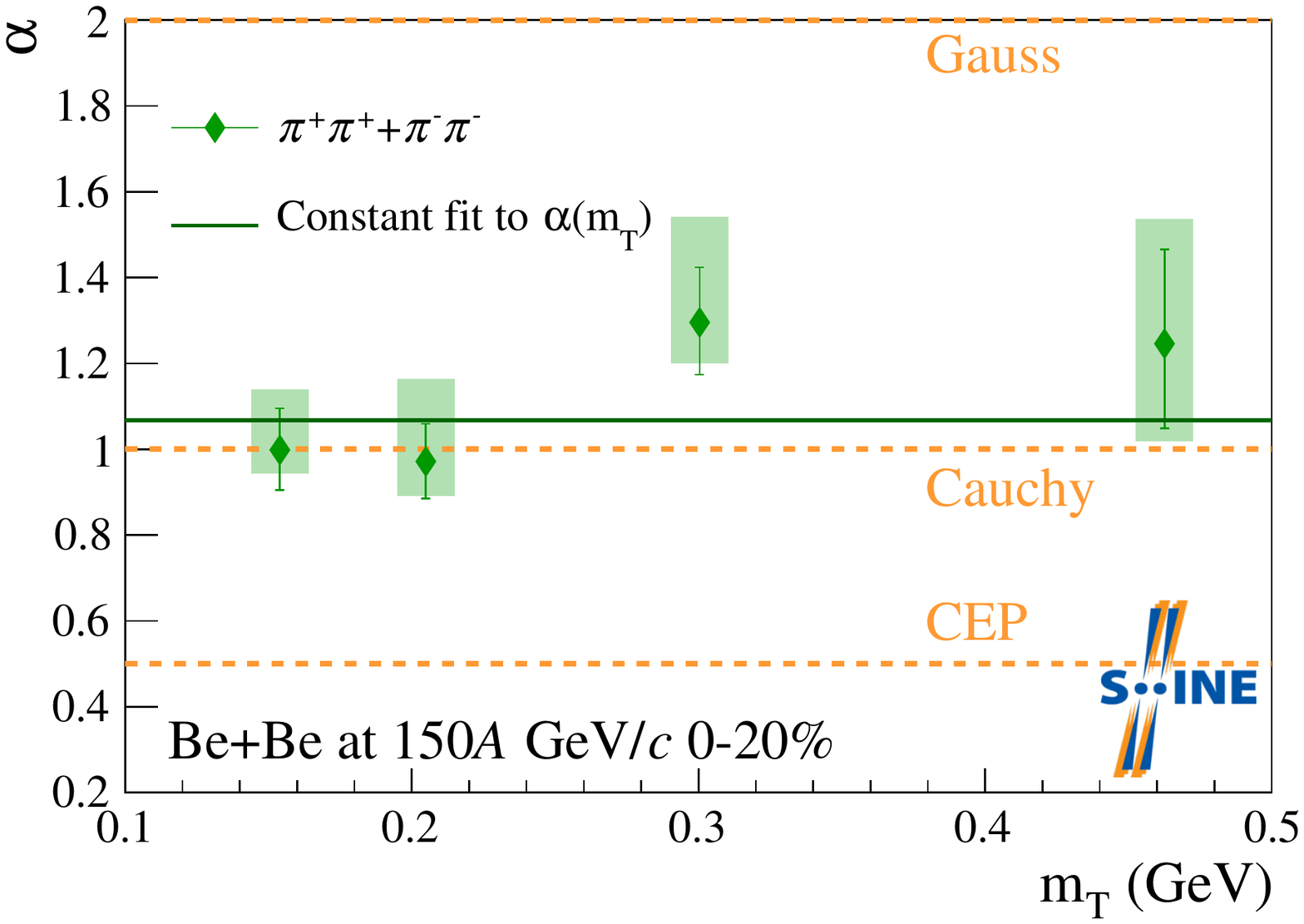}
     \caption{The L\'evy stability index $\alpha$, for 0--20\% central Be+Be at 150\AGeVc, as a function of \mt.
      Special cases corresponding to a Gaussian ($\alpha=2$) or a Cauchy ($\alpha=1$) source are shown, as well as $\alpha=0.5$, the conjectured value corresponding to the critical endpoint, while the constant $\alpha$ fit is shown with a solid line. Boxes denote systematic uncertainties, bars represent statistical ones.}
     \label{fig:alpha}
\end{figure}

The L\'evy stability exponent $\alpha$ describes the shape of the tail of the source
distribution. The \NASixtyOne results, shown in Fig.~\ref{fig:alpha}, yield values for $\alpha$ between 0.9 and 1.5, and are significantly lower than the Gaussian ($\alpha = 2$) case, and also significantly higher than the conjectured critical endpoint value ($\alpha = 0.5$). The obtained $\alpha$ values are in a similar range as the ones obtained in Au+Au collisions at RHIC energies~\cite{PHENIX:2017ino}.
The shape of the pion emitting source is apparently independent of \mt, within uncertainties. Therefore one can calculate a simple average of the four $\alpha$ values via a constant fit, shown in Fig.~\ref{fig:alpha}, and resulting in a good fit quality ($\chi^2/\textnormal{NDF}=6.0/2$, corresponding to a confidence level of 11\%). This results in an average value of \;$\overline \alpha = 1.07 \pm 0.06$ (stat.), which describes a source shape close to a Cauchy distribution (where $\alpha = 1$). Further studies are foreseen at \NASixtyOne using different collision energies and system sizes in order to map the evolution of the L\'evy stability index $\alpha$ as a function of collision energy and system size.

Furthermore $\alpha \ll 2$ is interesting in particular as we observe a $R \sim 1/\sqrt{\mt}$, visible in Fig.~\ref{fig:R}. It is not entirely clear why it is the case (the indicated \mt dependence could form in the QGP or at a later stage), as one would expect this to show only at $\alpha = 2$\cite{Sinyukov:1994vg}; this phenomenon was observed also at RHIC~\cite{PHENIX:2017ino}.

\section{Summary and conclusions}
\label{sec:Conclusion}
Measurement of two-particle femtoscopic correlations in 150\AGeVc Be+Be collisions with the 
\break \NASixtyOne detector system was presented. The correlation functions were measured in several bins of pair transverse mass \mt, and their fundamental shape parameters were extracted via fitting a L\'evy shaped ansatz for the particle source function. Correction for the final state Coulomb interaction was performed. The \mt-dependence of the shape parameters $\lambda$, $R$ and $\alpha$ were studied. 

The results show that the L\'evy exponent $\alpha$ is approximately constant as a function of \mt, and far from both the Gaussian case of $\alpha=2$ or the conjectured value at the critical endpoint, $\alpha=0.5$. The radius scale parameter $R$ shows a slight decrease in \mt, which can be explained as a signature of transverse flow. Finally, an approximately constant trend of the intercept parameter $\lambda$ as a function of \mt was observed, clearly different from measurement results at RHIC, but similar to previous NA44 measurements based on a Gaussian approximation. The \NASixtyOne experimental program plans further measurements at different energies and system sizes of these L\'evy shape parameters. This will complete a systematic study of the energy and system size dependence of the source shape parameters.

%***********************************************************************************

\appendix
\section{Additional L\'evy fits}
\begin{figure}[H]
\centering
\includegraphics[width=0.6\textwidth]{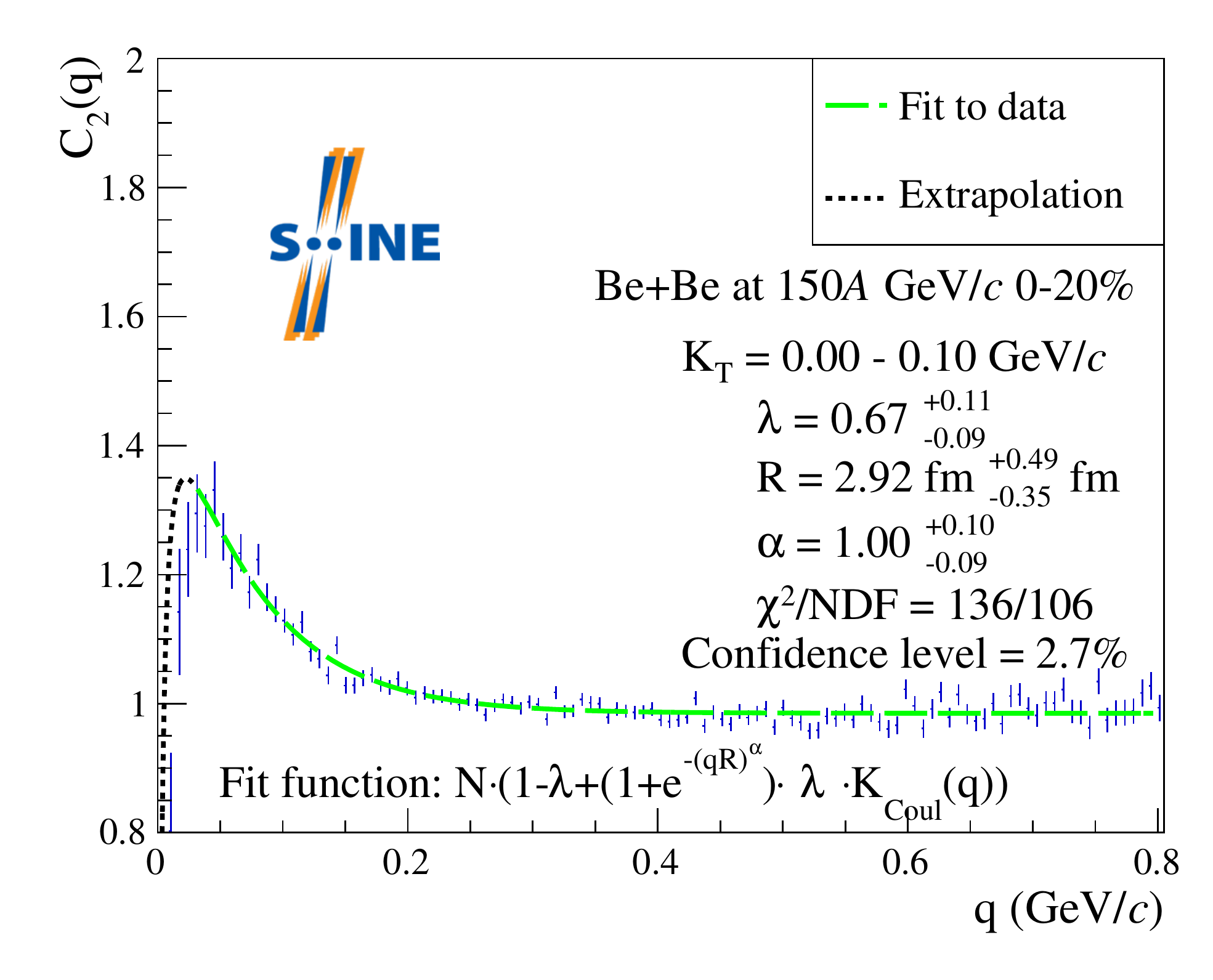}
\caption{Measured Bose-Einstein correlation function and the corresponding fit, similarly to Fig.~\ref{fig:examplefit}, but for $\kt = 0.00\--0.10$~\GeVc. Fit range in $q$ in this case was 0.028~\GeVc to 0.8~\GeVc.}
\label{fig:FitKT0}
\end{figure}

\begin{figure}[H]
\centering
\includegraphics[width=0.6\textwidth]{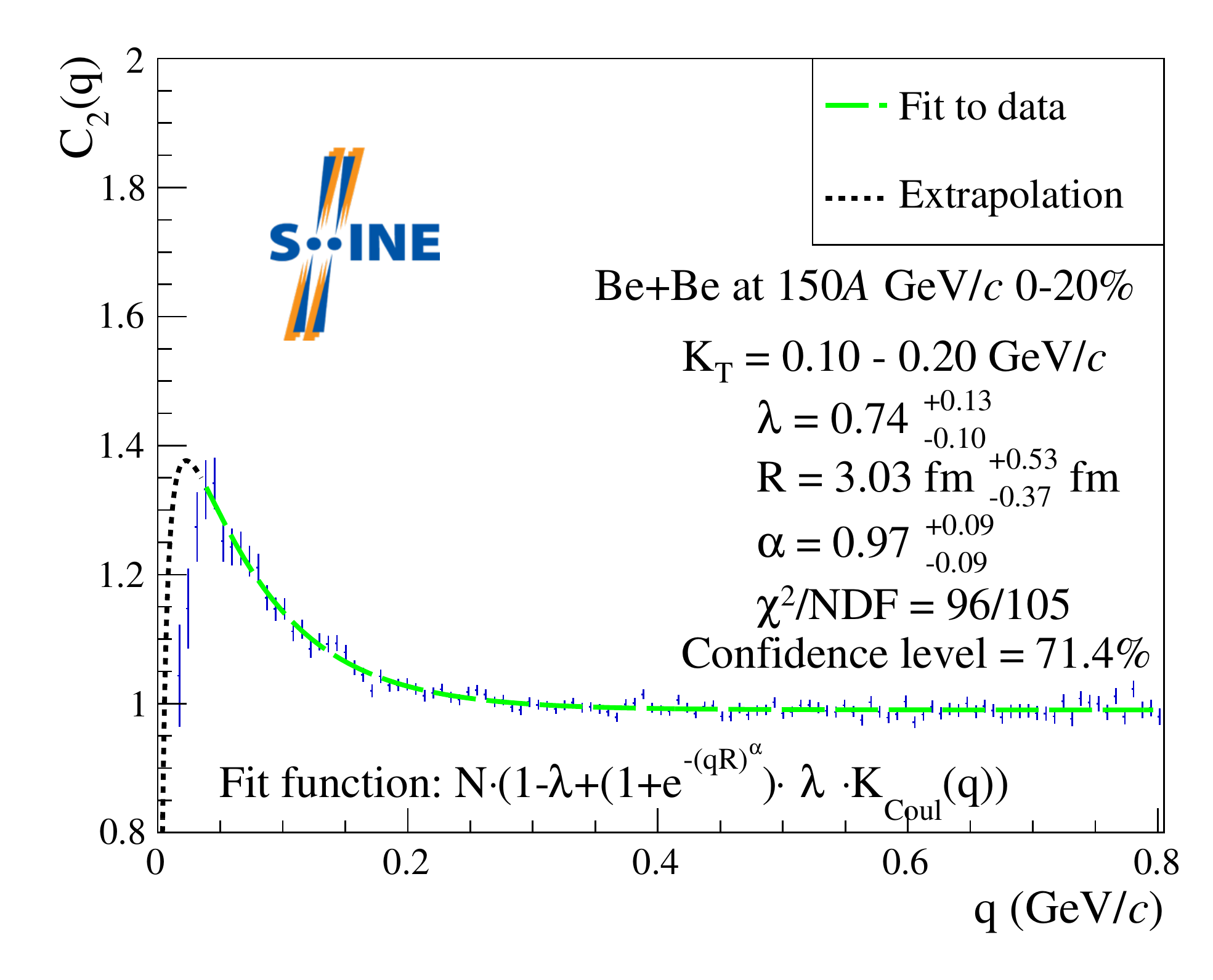}
\caption{Measured Bose-Einstein correlation function and the corresponding fit, similarly to Fig.~\ref{fig:examplefit}, but for $\kt = 0.10\--0.20$~\GeVc. Fit range in $q$ in this case was 0.035~\GeVc to 0.8~\GeVc.}
\label{fig:FitKT1}
\end{figure}

\begin{figure}[H]
\centering
\includegraphics[width=0.6\textwidth]{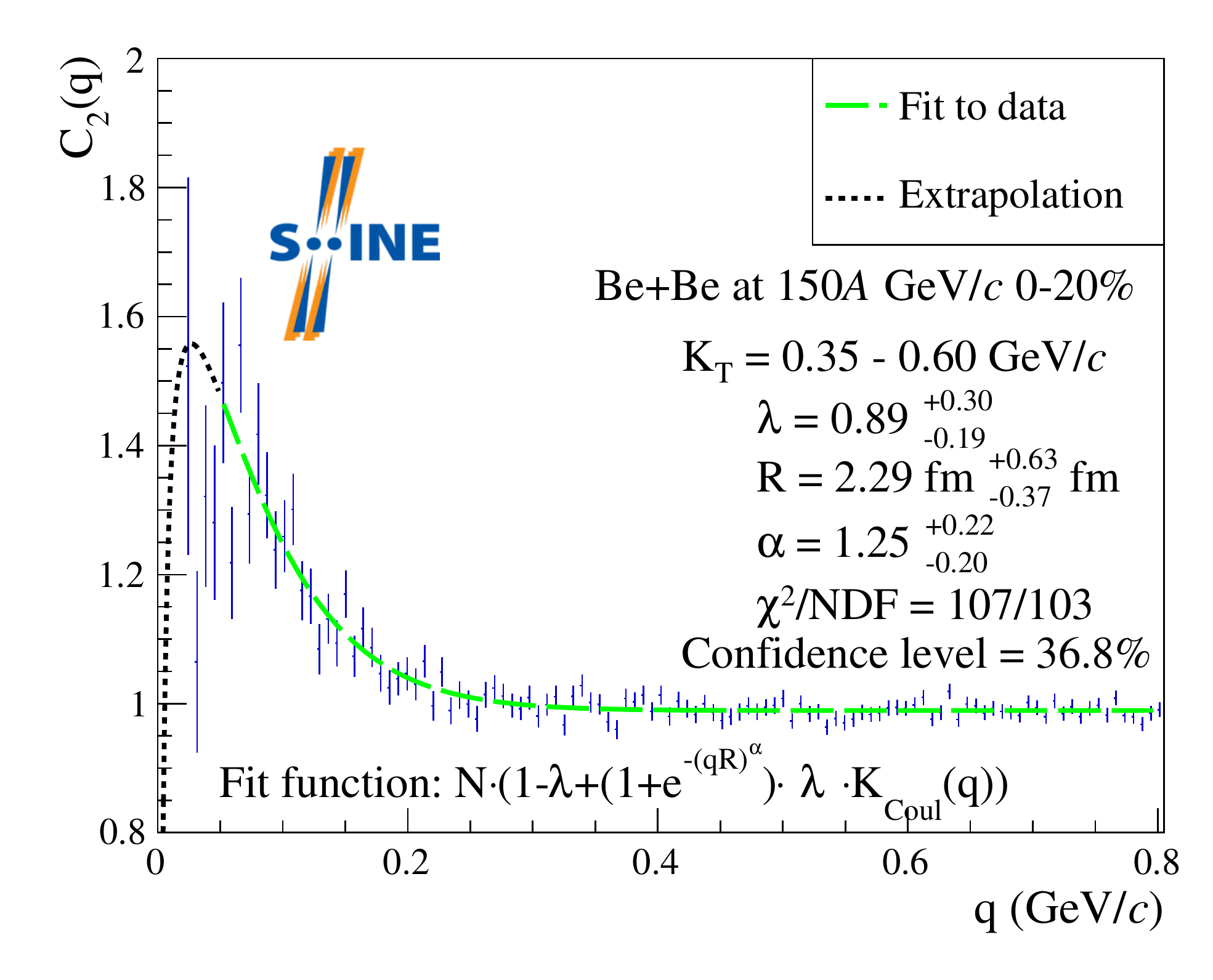}
\caption{Measured Bose-Einstein correlation function and the corresponding fit, similarly to Fig.~\ref{fig:examplefit}, but for $\kt = 0.35\--0.60$~\GeVc. Fit range in $q$ in this case was 0.049~\GeVc to 0.8~\GeVc.}
\label{fig:FitKT3}
\end{figure}

Figures \ref{fig:FitKT0}-\ref{fig:FitKT3} show the individual correlation function measurements and the corresponding fits in each of the $\kt$ intervals, separately.
 
\section{Systematic uncertainties}
 Tables \ref{tab:Rsyssources}-\ref{tab:Asyssources} contain the physical parameter values from L\'evy fits and their systematic uncertainty contributions from various sources. Values in the table indicate the change of the given fit parameter when the indicated analysis setting was modified, as compared to its default value. In these tables, an uncertainty value of $0.00\%$ means that the given setting modified the fit parameter value only in the opposite direction.

\begin{table}[H]
\begin{tabular}{c|cc|cc|cc|cc|}
\hline
\multicolumn{1}{|c|}{\kt range}    & \multicolumn{2}{c|}{$0.00-0.10$~\GeVc} & \multicolumn{2}{c|}{$0.10-0.20$~\GeVc}  & \multicolumn{2}{c|}{$0.20-0.35$~\GeVc} & \multicolumn{2}{c|}{$0.35-0.60$~\GeVc}  \\ \hline
                                     & \multicolumn{2}{c|}{$R = 2.92$ fm} 
                                     & \multicolumn{2}{c|}{$R = 3.03$ fm} 
                                     & \multicolumn{2}{c|}{$R = 2.18$ fm} 
                                     & \multicolumn{2}{c|}{$R = 2.29$ fm}       \\ \cline{2-9} 
                                     & \multicolumn{1}{c|}{Up}     & Down    & \multicolumn{1}{c|}{Up}      & Down    & \multicolumn{1}{c|}{Up}     & Down    & \multicolumn{1}{c|}{Up}      & Down    \\ \hline
\multicolumn{1}{|c|}{Fit range}      & \multicolumn{1}{c|}{0.00\%} & 0.07\%  & \multicolumn{1}{c|}{2.18\%}  & 0.00\%  & \multicolumn{1}{c|}{0.00\%} & 1.57\%  & \multicolumn{1}{c|}{0.00\%}  & 2.30\%  \\ \hline
\multicolumn{1}{|c|}{PID cut}        & \multicolumn{1}{c|}{1.15\%} & 0.00\%  & \multicolumn{1}{c|}{4.24\%}  & 0.00\%  & \multicolumn{1}{c|}{1.61\%} & 4.55\%  & \multicolumn{1}{c|}{28.00\%} & 0.00\%  \\ \hline
\multicolumn{1}{|c|}{VTPC points}    & \multicolumn{1}{c|}{0.61\%} & 2.61\%  & \multicolumn{1}{c|}{4.92\%}  & 0.00\%  & \multicolumn{1}{c|}{0.00\%} & 4.81\%  & \multicolumn{1}{c|}{8.03\%}  & 1.10\%  \\ \hline
\multicolumn{1}{|c|}{$|B_x|, |B_y|$} & \multicolumn{1}{c|}{0.00\%} & 5.83\%  & \multicolumn{1}{c|}{1.98\%}  & 6.75\%  & \multicolumn{1}{c|}{0.00\%} & 7.48\%  & \multicolumn{1}{c|}{4.50\%}  & 10.06\% \\ \hline
\multicolumn{1}{|c|}{nPoint}         & \multicolumn{1}{c|}{9.23\%} & 0.00\%  & \multicolumn{1}{c|}{14.59\%} & 14.26\% & \multicolumn{1}{c|}{6.41\%} & 10.46\% & \multicolumn{1}{c|}{7.20\%}  & 15.61\% \\ \hline
\multicolumn{1}{|c|}{nPointRatio}    & \multicolumn{1}{c|}{0.94\%} & 13.78\% & \multicolumn{1}{c|}{3.56\%}  & 11.56\% & \multicolumn{1}{c|}{2.37\%} & 6.38\%  & \multicolumn{1}{c|}{6.62\%}  & 8.94\%  \\ \hline
\multicolumn{1}{|c|}{$q$ bin width}  & \multicolumn{1}{c|}{0.00\%} & 1.28\%  & \multicolumn{1}{c|}{0.00\%}  & 1.92\%  & \multicolumn{1}{c|}{2.73\%} & 0.00\%  & \multicolumn{1}{c|}{0.00\%}  & 4.97\%  \\ \hline
\multicolumn{1}{|c|}{vertex $z$}     & \multicolumn{1}{c|}{7.38\%} & 7.77\%  & \multicolumn{1}{c|}{0.00\%}  & 7.04\%  & \multicolumn{1}{c|}{0.00\%} & 8.57\%  & \multicolumn{1}{c|}{0.00\%}  & 12.76\% \\ \hline
\end{tabular}
\caption{Systematic uncertainties for L\'evy scale parameter $R$.}
\label{tab:Rsyssources}
\end{table}

\begin{table}[H]
\begin{tabular}{c|cc|cc|cc|cc|}
\hline
\multicolumn{1}{|c|}{\kt range}    & \multicolumn{2}{c|}{$0.00-0.10$~\GeVc} & \multicolumn{2}{c|}{$0.10-0.20$~\GeVc}  & \multicolumn{2}{c|}{$0.20-0.35$~\GeVc}  & \multicolumn{2}{c|}{$0.35-0.60$~\GeVc}  \\ \hline
                                     & \multicolumn{2}{c|}{$\lambda = 0.67$} & \multicolumn{2}{c|}{$\lambda = 0.74$}  & \multicolumn{2}{c|}{$\lambda = 0.63$}  & \multicolumn{2}{c|}{$\lambda = 0.89$}  \\ \cline{2-9} 
                                     & \multicolumn{1}{c|}{Up}     & Down    & \multicolumn{1}{c|}{Up}      & Down    & \multicolumn{1}{c|}{Up}      & Down    & \multicolumn{1}{c|}{Up}      & Down    \\ \hline
\multicolumn{1}{|c|}{Fit range}      & \multicolumn{1}{c|}{0.05\%} & 0.00\%  & \multicolumn{1}{c|}{2.39\%}  & 0.00\%  & \multicolumn{1}{c|}{0.00\%}  & 1.96\%  & \multicolumn{1}{c|}{0.00\%}  & 2.68\%  \\ \hline
\multicolumn{1}{|c|}{PID cut}        & \multicolumn{1}{c|}{0.99\%} & 1.56\%  & \multicolumn{1}{c|}{4.73\%}  & 0.00\%  & \multicolumn{1}{c|}{0.15\%}  & 0.25\%  & \multicolumn{1}{c|}{41.13\%} & 0.00\%  \\ \hline
\multicolumn{1}{|c|}{VTPC points}    & \multicolumn{1}{c|}{1.15\%} & 4.41\%  & \multicolumn{1}{c|}{3.75\%}  & 0.00\%  & \multicolumn{1}{c|}{0.01\%}  & 10.11\% & \multicolumn{1}{c|}{7.54\%}  & 0.88\%  \\ \hline
\multicolumn{1}{|c|}{$|B_x|, |B_y|$} & \multicolumn{1}{c|}{0.19\%} & 3.33\%  & \multicolumn{1}{c|}{2.00\%}  & 3.21\%  & \multicolumn{1}{c|}{0.00\%}  & 8.74\%  & \multicolumn{1}{c|}{5.77\%}  & 15.10\% \\ \hline
\multicolumn{1}{|c|}{nPoint}         & \multicolumn{1}{c|}{6.34\%} & 0.00\%  & \multicolumn{1}{c|}{22.77\%} & 15.61\% & \multicolumn{1}{c|}{10.41\%} & 13.19\% & \multicolumn{1}{c|}{5.51\%}  & 21.65\% \\ \hline
\multicolumn{1}{|c|}{nPointRatio}    & \multicolumn{1}{c|}{0.19\%} & 15.15\% & \multicolumn{1}{c|}{1.01\%}  & 16.88\% & \multicolumn{1}{c|}{0.00\%}  & 7.79\%  & \multicolumn{1}{c|}{5.96\%}  & 16.77\% \\ \hline
\multicolumn{1}{|c|}{$q$ bin width}  & \multicolumn{1}{c|}{0.00\%} & 1.25\%  & \multicolumn{1}{c|}{0.00\%}  & 1.99\%  & \multicolumn{1}{c|}{3.85\%}  & 0.00\%  & \multicolumn{1}{c|}{0.00\%}  & 6.38\%  \\ \hline
\multicolumn{1}{|c|}{vertex $z$}     & \multicolumn{1}{c|}{8.35\%} & 8.74\%  & \multicolumn{1}{c|}{0.05\%}  & 11.26\% & \multicolumn{1}{c|}{0.00\%}  & 12.10\% & \multicolumn{1}{c|}{0.00\%}  & 17.46\% \\ \hline
\end{tabular}
\caption{Systematic uncertainties for correlation strength parameter $\lambda$.}
\label{tab:Lsyssources}
\end{table}

\begin{table}[H]
\begin{tabular}{c|ll|ll|ll|ll|}
\hline
\multicolumn{1}{|c|}{\kt range}    & \multicolumn{2}{c|}{$0.00-0.10$~\GeVc}                   & \multicolumn{2}{c|}{$0.10-0.20$~\GeVc}                   & \multicolumn{2}{c|}{$0.20-0.35$~\GeVc}                   & \multicolumn{2}{c|}{$0.35-0.60$~\GeVc}                   \\ \hline
                                     & \multicolumn{2}{c|}{$\alpha = 1.00$}                     & \multicolumn{2}{c|}{$\alpha = 0.97$}                     & \multicolumn{2}{c|}{$\alpha = 1.30$}                     & \multicolumn{2}{c|}{$\alpha = 1.25$}                     \\ \cline{2-9} 
                                     & \multicolumn{1}{c|}{Up}      & \multicolumn{1}{c|}{Down} & \multicolumn{1}{c|}{Up}      & \multicolumn{1}{c|}{Down} & \multicolumn{1}{c|}{Up}      & \multicolumn{1}{c|}{Down} & \multicolumn{1}{c|}{Up}      & \multicolumn{1}{c|}{Down} \\ \hline
\multicolumn{1}{|c|}{Fit range}      & \multicolumn{1}{l|}{0.00\%}  & 0.16\%                    & \multicolumn{1}{l|}{0.00\%}  & 1.44\%                    & \multicolumn{1}{l|}{0.00\%}  & 0.00\%                    & \multicolumn{1}{l|}{0.00\%}  & 0.00\%                    \\ \hline
\multicolumn{1}{|c|}{PID cut}        & \multicolumn{1}{l|}{0.00\%}  & 0.99\%                    & \multicolumn{1}{l|}{2.76\%}  & 1.97\%                    & \multicolumn{1}{l|}{0.00\%}  & 2.71\%                    & \multicolumn{1}{l|}{0.00\%}  & 13.77\%                   \\ \hline
\multicolumn{1}{|c|}{VTPC points}    & \multicolumn{1}{l|}{2.09\%}  & 0.60\%                    & \multicolumn{1}{l|}{2.76\%}  & 2.52\%                    & \multicolumn{1}{l|}{2.09\%}  & 0.12\%                    & \multicolumn{1}{l|}{2.09\%}  & 6.66\%                    \\ \hline
\multicolumn{1}{|c|}{$|B_x|, |B_y|$} & \multicolumn{1}{l|}{8.55\%}  & 0.33\%                    & \multicolumn{1}{l|}{6.66\%}  & 1.30\%                    & \multicolumn{1}{l|}{8.55\%}  & 0.00\%                    & \multicolumn{1}{l|}{8.55\%}  & 2.90\%                    \\ \hline
\multicolumn{1}{|c|}{nPoint}         & \multicolumn{1}{l|}{0.00\%}  & 3.74\%                    & \multicolumn{1}{l|}{20.24\%} & 7.17\%                    & \multicolumn{1}{l|}{0.00\%}  & 5.63\%                    & \multicolumn{1}{l|}{0.00\%}  & 6.90\%                    \\ \hline
\multicolumn{1}{|c|}{nPointRatio}    & \multicolumn{1}{l|}{10.18\%} & 0.29\%                    & \multicolumn{1}{l|}{9.96\%}  & 1.86\%                    & \multicolumn{1}{l|}{10.18\%} & 3.26\%                    & \multicolumn{1}{l|}{10.18\%} & 6.68\%                    \\ \hline
\multicolumn{1}{|c|}{$q$ bin width}  & \multicolumn{1}{l|}{0.92\%}  & 0.00\%                    & \multicolumn{1}{l|}{3.77\%}  & 0.00\%                    & \multicolumn{1}{l|}{0.92\%}  & 1.94\%                    & \multicolumn{1}{l|}{0.92\%}  & 0.00\%                    \\ \hline
\multicolumn{1}{|c|}{vertex $z$}     & \multicolumn{1}{l|}{4.36\%}  & 3.76\%                    & \multicolumn{1}{l|}{6.92\%}  & 0.00\%                    & \multicolumn{1}{l|}{4.36\%}  & 0.00\%                    & \multicolumn{1}{l|}{4.36\%}  & 0.00\%                    \\ \hline
\end{tabular}
\caption{Systematic uncertainties for L\'evy stability index $\alpha$.}
\label{tab:Asyssources}
\end{table}

%***********************************************************************************
\section*{Acknowledgments}
%{\red{ Please check with the most recent official \NASixtyOne acknowledgments as well.}}
We would like to thank the CERN EP, BE, HSE and EN Departments for the
strong support of NA61/SHINE.

This work was supported by
the Hungarian Scientific Research Fund (grant NKFIH 138136\slash138152 and TKP2021-NKTA-64),
the Polish Ministry of Science and Higher Education
(DIR\slash WK\slash\-2016\slash 2017\slash\-10-1, WUT ID-UB), the National Science Centre Poland (grants
2014\slash 14\slash E\slash ST2\slash 00018, %AR, settled
2016\slash 21\slash D\slash ST2\slash 01983, %MMP, settled
2017\slash 25\slash N\slash ST2\slash 02575, %AT, settled
2018\slash 29\slash N\slash ST2\slash 02595, %AM, completed, not settled
2018\slash 30\slash A\slash ST2\slash 00226, %MG, in progress
2018\slash 31\slash G\slash ST2\slash 03910, %SK, in progress
2019\slash 33\slash B\slash ST9\slash 03059, %DB, LT (astro), in progress
2020\slash 39\slash O\slash ST2\slash 00277), %MR, in progress
the Norwegian Financial Mechanism 2014--2021 (grant 2019\slash 34\slash H\slash ST2\slash 00585),
the Polish Minister of Education and Science (contract No. 2021\slash WK\slash 10),
the Russian Science Foundation (grant 17-72-20045),
the Russian Academy of Science and the
Russian Foundation for Basic Research (grants 08-02-00018, 09-02-00664
and 12-02-91503-CERN),
the Russian Foundation for Basic Research (RFBR) funding within the research project no. 18-02-40086,
the Ministry of Science and Higher Education of the Russian Federation, Project "Fundamental properties of elementary particles and cosmology" No 0723-2020-0041,
the European Union's Horizon 2020 research and innovation programme under grant agreement No. 871072,
the Ministry of Education, Culture, Sports,
Science and Tech\-no\-lo\-gy, Japan, Grant-in-Aid for Sci\-en\-ti\-fic
Research (grants 18071005, 19034011, 19740162, 20740160 and 20039012),
the German Research Foundation DFG (grants GA\,1480\slash8-1 and project 426579465),
the Bulgarian Ministry of Education and Science within the National
Roadmap for Research Infrastructures 2020--2027, contract No. D01-374/18.12.2020,
Ministry of Education
and Science of the Republic of Serbia (grant OI171002), Swiss
Nationalfonds Foundation (grant 200020\-117913/1), ETH Research Grant
TH-01\,07-3 and the Fermi National Accelerator Laboratory (Fermilab), a U.S. Department of Energy, Office of Science, HEP User Facility managed by Fermi Research Alliance, LLC (FRA), acting under Contract No. DE-AC02-07CH11359 and the IN2P3-CNRS (France).\\

The data used in this paper were collected before February 2022.

\bibliographystyle{na61Utphys}
%{\footnotesize\raggedright
\bibliography{na61Preprint.bib}
%}
\newpage
{\Large The \NASixtyOne Collaboration}
\bigskip
\begin{sloppypar}
% based on XML DB with time Thu Jan 26 09:03:08 2023
% Authors in alphabetical order.

\noindent
H.~Adhikary$^{\,13}$,
P.~Adrich$^{\,15}$,
K.K.~Allison$^{\,29}$,
N.~Amin$^{\,5}$,
E.V.~Andronov$^{\,25}$,
T.~Anti\'ci\'c$^{\,3}$,
I.-C.~Arsene$^{\,12}$,
M.~Bajda$^{\,16}$,
Y.~Balkova$^{\,18}$,
M.~Baszczyk$^{\,17}$
% ...

\end{sloppypar}
% based on XML DB with time Thu Jan 26 09:03:08 2023
% Institutes in alphabetical order.

\noindent
% ...
$^{3}$~Ru{\dj}er Bo\v{s}kovi\'c Institute, Zagreb, Croatia\\
% ...
$^{5}$~Karlsruhe Institute of Technology, Karlsruhe, Germany\\
% ...
$^{12}$~University of Oslo, Oslo, Norway\\
$^{13}$~Jan Kochanowski University, Poland\\
% ...
$^{15}$~National Centre for Nuclear Research, Warsaw, Poland\\
$^{16}$~Jagiellonian University, Cracow, Poland\\
$^{17}$~AGH - University of Science and Technology, Cracow, Poland\\
$^{18}$~University of Silesia, Katowice, Poland\\
% ...
$^{25}$~St. Petersburg State University, St. Petersburg, Russia\\
% ...
$^{29}$~University of Colorado, Boulder, USA\\
% ...

\end{document}